\documentclass[a4paper,graphics,natbib,graphicx,psfig,amssymb,ccaption]{article}
\usepackage{lscape}
\usepackage{graphicx}
\usepackage{subcaption}
\newcommand{\affil}[1]{$^{\rm #1}$}
\usepackage[authoryear]{natbib}
\bibpunct{(}{)}{;}{a}{}{,}
\date{} 

\title{\large\bf\flushleft Absolute Magnitude Calibration for Dwarfs Based on the Colour-Magnitude Diagrams of Galactic Clusters}
\author{\parbox{\textwidth}{\flushleft
\vspace{-0.5cm}
{\it S. Karaali\affil{\dag, A, B}, E. Yaz G\"ok\c ce\affil{A}, S. Bilir\affil{A}, and S. Tun\c cel G\"u\c ctekin\affil{A}}\\
\vspace{0.4cm}
{\small \affil{A}\,Istanbul University, Faculty of Science, Department of Astronomy and Space Sciences, 34119, Istanbul, Turkey}\\
{\small \affil{B}\,Email: karsa@istanbul.edu.tr}}}
\begin{document}
\twocolumn[
\begin{changemargin}{.8cm}{.5cm}
\begin{minipage}{.9\textwidth}
\vspace{-1cm}
\maketitle
\small{\bf Abstract:}
We present two absolute magnitude calibrations for dwarfs based on colour-magnitude 
diagrams of Galactic clusters. The combination of the $M_g$ absolute magnitudes of 
the dwarf fiducial sequences of the clusters M92, M13, M5, NGC 2420, M67 and NGC 6791 
with the corresponding metallicities provides absolute magnitude calibration for a 
given $(g-r)_0$ colour. The calibration is defined in the colour interval 
$0.25\leq (g-r)_0 \leq 1.25$ mag and it covers the metallicity interval 
$-2.15\leq \lbrack Fe/H\rbrack \leq +0.37$ dex. The absolute magnitude residuals obtained 
by the application of the procedure to another set of Galactic clusters lie in the interval 
$-0.15 \leq \Delta M_g \leq +0.12$ mag. The mean and standard deviation of the residuals 
are $<\Delta M_g>=-0.002$ and $\sigma=0.065$ mag, respectively. The calibration of the 
$M_J$ absolute magnitude in terms of metallicity is carried out 
by using the fiducial sequences of the clusters M92, M13, 47 Tuc, NGC 2158 and NGC 6791. 
It is defined in the colour interval $0.90 \leq (V-J)_0 \leq 1.75$ mag and it covers the 
same metallicity interval of the $M_g$ calibration. The absolute magnitude residuals 
obtained by the application of the procedure to the cluster M5 ($\lbrack Fe/H\rbrack=-1.40$ dex) 
and 46 solar metallicity, $-0.45 \leq \lbrack Fe/H\rbrack \leq +0.35$ dex, field stars 
lie in the interval $-0.29$ and $+0.35$ mag. However, the range of 87 per cent of them is 
rather shorter, $-0.2 \leq \Delta M_J \leq +0.2$ mag. The mean and standard deviation 
of all residuals are $<\Delta M_J>=0.05$ and $\sigma=0.13$ mag, respectively. 
The derived relations are applicable to stars older than 4 Gyr for the $M_g$ calibration, 
and older than 2 Gyr for the $M_J$ calibration. The cited limits are the ages of the youngest 
calibration clusters in the two systems.
    
\medskip
\medskip
\end{minipage}
\end{changemargin}
]
\small
\let\thefootnote\relax\footnote{\small \affil{\dag}\,Retired.}
\section{Introduction}
The distance of an astronomical object plays an important role in deriving absolute magnitudes of 
stars and determining the three-dimensional structure of our Galaxy. For nearby stars, 
{\em Hipparcos} \citep{ESA97} is the main supplier of the data where a trigonometric parallax has 
been used. However, for stars at large distances a photometric parallax is inevitable. 

Different methods appeared in the literature related to absolute magnitude determination. \cite{Nissen91} 
and \cite{Laird88} used the offset from standard main-sequence for the absolute magnitude determination of 
dwarfs. The studies of \cite{Phleps00} and \cite{C01} are based on the colour-absolute magnitude 
diagrams of some specific clusters whose metal abundances are adopted as representative of a Galactic 
population, i.e. thin and thick discs, and halo. \cite{Siegel02} derived two relations, one for solar-like
abundances and another for metal-poor stars, between the $M_R$ absolute magnitude and $R-I$ colour 
index which provide absolute magnitude estimation of dwarfs with different metallicities by means of 
linear interpolation or extrapolation.  

A more recent procedure for absolute magnitude estimation for dwarfs and giants simultaneously is based on 
the measured atmospheric parameters and the time spent by the star in each region of the H-R diagram. 
For details of this procedure, we refer to the works of \cite{Breddels10} and \cite{Zwitter10}.

The most recent absolute magnitude calibrations are those of \citet*{Karaali12,Karaali13a,Karaali13b}. 
Karaali et al. combined the colour-apparent magnitude diagrams of Galactic clusters with different 
metal abundances and their distance moduli, and calibrated the $M_V$, $M_g$, $M_J$, and $M_{K_s}$ 
absolute magnitudes of red giants in terms of metallicity. Here, we will apply the same procedure to 
the dwarfs using the Sloan Digital Sky Survey \citep[SDSS;][]{York00} and Two Micron All Sky Survey 
\citep[2MASS;][]{Skrutskie06} data, two of the widely used photometric systems in the Galactic researches. 
This procedure is different than the ones in \cite{Karaali03,Karaali05} where the absolute magnitude was calibrated 
in terms of the parameter $\delta_{0.6}$, the ultraviolet excess relative to the Hyades cluster, reduced to 
the intrinsic colour $B-V=0.60$ mag. The $U$ ultraviolet magnitude can not be observed accurately for each star. 
Whereas, various procedures can be utilized to determine metallicity for a larger set of stars. Hence, 
for the procedure in this paper we expect an advantage over the former one. 

This paper is the fourth one related to the calibration of absolute magnitudes of stars. The former ones  
were devoted to red giants, while in this paper the subject is dwarfs. We calibrate the $M_g$ and $M_J$ 
absolute magnitudes in terms of metallicity. The outline of the paper follows the presentation of the 
data in Section 2, procedure utilized for calibration in Section 3, and a summary and discussion in Section 4. 

\section{Data}
We calibrated two different absolute magnitudes, $M_g$ and $M_J$, in terms of metallicity. Hence we used 
two different sets of data. The calibration of $g_0$ with $(g-r)_0$ is given in Section 2.1, while that 
for $J_0$ with $(V-J)_0$ is presented in Section 2.2.	

\subsection{Data for Calibration with SDSS: $g_0 \times (g-r)_0$}
Our sample consists of M92, M13, M5, NGC 2420, M67, and NGC 6791 stellar clusters with different metal abundance. 
The SDSS photometric data of the clusters were taken from \cite{An08}, while the $(g-M_g)_0$ 
true distance modulus, $E(B-V)$ colour excess, and $\lbrack Fe/H\rbrack$ iron abundance are from 
the references in the last column of Table 1. The range of the metallicity is $-2.15\leq \lbrack Fe/H\rbrack \leq
+0.37$ dex. The $g$ and $g-r$ fiducials of the stellar clusters adapted from \cite{An08} are given in Table 2. 
As in our previous works, we used $R_V=3.1$ \citep*{Cardelli89} to obtain the total extinction at the $V$ band. Then, we used 
the equations of \cite{Fan99} to calculate the total extinction in $A_g$ and the selective absorption in SDSS, 
$E(g-r)/A_{V}=0.341$. 

We fitted the $g_0\times(g-r)_0$ main-sequence of each cluster to a fifth degree polynomial. 
The calibration of $g_0$ is as follows:

\begin{eqnarray}
g_0=\sum_{i=0}^{5}a_i(g-r)_0^{i}
\end{eqnarray}
The numerical values of $a_i$ coefficients are given in Table 3 and the corresponding diagrams are shown 
in Fig. 1. The $(g-r)_0$ values in the second line of the table denote the range of $(g-r)_0$ 
available for each cluster. 

\subsection{Data for Calibration with 2MASS: $J_0 \times (V-J)_0$}
Five clusters with different metallicities, i.e. M92, M13, NGC 6791, 47 Tuc and NGC 2158, 
were selected for our programme. The $J$ and $V-J$ data for the first three clusters were taken 
from \citet{Brasseur10}, while those for the clusters of 47 Tuc and NGC 2158 were transformed 
from the $BVRI$ and $gri$ data, respectively, as explained in the following. For 47 Tuc, 
we de-reddened the $V$, $B-V$, and $R-I$ magnitude and colours of \citet{Alcaino90} and 
transformed them to the $(V-J)_0$ colours by the following equation of \citet{Bilir08}:
\begin{equation}
(V-J)_0=1.557(B-V)_0+0.461(R-I)_0+0.049.               
\end{equation}
Then, we combined the $V_0$ magnitude and $(V-J)_0$ colour and obtained the $J_0$ magnitude. 
For the cluster NGC 2158, we obtained the $(V-J)_0$ colours in two steps. First, we transformed 
the de-reddened $g_0$, $(g-r)_0$, and $(r-i)_0$ data of \citet{Smolinski11} to the $(g-J)_0$ 
colour by using the following equation of \citet{Bilir08}:
\begin{equation}
(g-J)_0=1.536(g-r)_0+1.400(r-i)_0+0.488.	
\end{equation}
Then, the magnitude $g_0$ is combined with this equation to give the $J_0$ magnitude, and further, Eq. (2) of \citet{Chonis08} given in the following:      

\begin{equation}
V_0=g_0-0.587(g-r)_0 -0.011,                                               
\end{equation}
is used in the evaluation of $(V-J)_0$ as follows:
\begin{equation}
(V-J)_0=(g-J)_0-0.587(g-r)_0-0.011.  
\end{equation}
The range of the metallicity of the clusters given in iron abundance is $-2.15\leq \lbrack Fe/H \rbrack\leq 0.37$ dex. 
The true distance modulus, ${\mu_0}$, $E(B-V)$ colour excess and ${[Fe/H]}$ iron abundance for the clusters 
are taken from the authors given in the last column of Table 4. Two references are given for the clusters 
47 Tuc and NGC 2158. The metallicity of 47 Tuc is taken from \citet{Alcaino90}, while the colour excess 
and true distance modulus are from \citet{Harris96}. The metallicity and the colour excess of the cluster 
NGC 2158 are those of \citet{Smolinski11}, while its distance modulus is reduced from the apparent distance 
modulus in \citet{Arp62}, i.e. $\mu=14.74$ mag. Although this distance modulus is old, it provides $J_0$ and 
$(V-J)_0$ data which are more agreeable with the colours and magnitudes of the other clusters used in the 
calibration. The $J_0$ and $(V-J)_0$ data are presented in Table 5. We adopted the $R=A_V/E(B-V)=3.1$ to 
convert the colour excess to the extinction \citep{Cardelli89}. Then, we used the equations $A_J/E(B-V)=0.87$ and $E(V-J)/E(B-V)=2.25$ \citep{McCall04} to de-redden the $J$ magnitude and $V-J$ colour, respectively. The de-reddening of the $R-I$ colour used for the cluster 47 Tuc is carried out by the equation $E(R-I)/E(B-V)=0.60$ \citep{Bilir08}.

We fitted the $J_0\times (V-J)_0$ sequence of four clusters to a fifth degree polynomial, while a third 
degree polynomial was sufficient for NGC 2158 due to less data. The calibration of $J_0$ is as follows:
\begin{eqnarray}
J_{0}= \sum_{i=0}^{5}b_{i}(V-J)^{i}_0 
\end{eqnarray}
The numerical values of the coefficients $b_i$ $(i=0, 1, 2, 3, 4, 5)$ are given in Table 6 and the corresponding 
diagrams are presented in Fig. 2. The $(V-J)_0$ interval in the second line of the table denotes the range of 
$(V-J)_0$ available for each cluster. 
 
\section{The Procedure}
\subsection{Absolute Magnitude as a Function of Metallicity}
We adopted the procedure used in our previous papers \citep{Karaali13a,Karaali13b} 
which consists of calibration of an absolute magnitude as a function of metallicity. 
We calibrated the $M_g$ and $M_J$ absolute magnitudes in terms of metallicity for 
a given $(g-r)_0$ and $(V-J)_0$ colour, respectively.

\subsubsection{Calibration of $M_g$ in terms of Metallicity}
We estimated the $M_g$ absolute magnitude for the $(g-r)_0$ colours given in Table 7 
for the cluster sample in Table 1 by combining the $g_0$ apparent magnitude evaluated 
by Eq. (1) and the true distance modulus ($\mu_0$) of the cluster in question, i.e.

\begin{eqnarray}
M_g=g_0-\mu_0.                     
\end{eqnarray} 
The $M_g \times (g-r)_0$ absolute magnitude-colour diagrams are plotted in Fig. 3. The 
absolute magnitudes of the clusters M13 ($\lbrack Fe/H\rbrack=-1.41$ dex) and M5 
($\lbrack Fe/H\rbrack=-1.26$ dex) are close to each other for a given colour. Similar 
case is valid for the clusters NGC 2420 ($\lbrack Fe/H\rbrack=-0.37$ dex) and M67 
($\lbrack Fe/H\rbrack=-0.04$ dex). However, we used both couples to obtain a better 
absolute magnitude calibration.

In this phase, $M_g$ absolute magnitudes can be fitted to the corresponding 
$\lbrack Fe/H\rbrack$ metal abundance for a given $(g-r)_0$ colour index 
and obtain the required calibration. This procedure is executed for the colour 
indices $(g-r)_0=$0.25, 0.40, 0.65, 0.80, 0.95, and 1.20 (Table 8 and Fig. 4).
The absolute magnitudes in all colour indices are fitted to a second degree 
polynomial with (squared) correlation coefficients between $R^2=0.962$ and 
$R^2=0.999$, where the first one corresponds to the absolute magnitudes for the 
colour index $(g-r)_0=0.80$ mag (but see the following paragraph). A third degree 
polynomial (Fig. 5) increases the correlation coefficient to $R^2=$ 1. However, it causes a flat 
distribution in the metallicity interval $-0.90\leq \lbrack Fe/H\rbrack\leq-0.40$ dex, 
resulting almost a constant absolute magnitude in this metallicity interval which contradicts 
the trends of absolute magnitudes for small colour indices, such as $(g-r)_0=0.35$ mag, 
where the correlation coefficient is high, $R^2=0.991$. We shall see in Section 3.2 
that the application of the third degree polynomial in question causes larger residuals 
than the ones for a quadratic polynomial, $\Delta M=-0.360$ and 0.067 mag, respectively. 

The number of clusters in the absolute magnitude calibration for the colour indices 
$0.82 \leq(g-r)_{0}\leq 1.25$ mag is only three, i.e. NGC 2420, M67, and NGC 6791 
with metallicities $\lbrack Fe/H\rbrack=-0.37$, $\lbrack Fe/H\rbrack=-0.04$, 
and $\lbrack Fe/H\rbrack=0.37$ dex, respectively. Additionally, the absolute 
magnitudes of the clusters NGC 2420 and M67 are close to each other. If we fit a 
quadratic polynomial to three metallicity and absolute magnitude couples, the segment 
corresponding to the interval $-0.37 \leq \lbrack Fe/H\rbrack\leq-0.04$ dex will 
perform a concave shape resulting in (estimated) absolute magnitudes larger than the ones 
for the cluster NGC 2420 which is opposite to the sense of the argument used in our work, 
i.e. the absolute magnitude of a dwarf with a given metallicity and colour should be 
brighter than one relatively more metal poor. Hence, we fitted a linear equation 
to three metallicity and absolute magnitude couples in question (Fig. 6b).  

The procedure can be applied to any $(g-r)_0$ colour interval for which the sample 
clusters are defined. The $(g-r)_0$ domains of the clusters are different. Hence, 
we adopted this interval in our study as 0.25 $\leq(g-r)_{0}\leq$ 1.25 mag where 
at least three clusters are defined, and we evaluated $M_g$ absolute magnitudes 
for each colour. Then, we combined them with the corresponding $\lbrack Fe/H\rbrack$ 
metallicities and obtained the final calibrations. The general form of the equation 
for the calibrations is as follows:

\begin{eqnarray}
M_g = c_0 + c_1 X +c_2 X^2,
\end{eqnarray} 
where $X = \lbrack Fe/H\rbrack$. The absolute magnitudes estimated via Eq. (7) 
for 101 $(g-r)_0$ colour indices and the corresponding $c_i$ ($i=$ 0, 1, 2) coefficients 
are given in Table 9. The coefficients $c_1$ and $c_0$ for the linear equation    
\begin{eqnarray}
M_g = c_0 + c_1 X,
\end{eqnarray} 
which are valid for the colour interval $0.82\leq(g-r)_{0}\leq1.25$ are also tabulated 
in Table 9. The diagrams for the calibrations are not given in the paper because of 
space constraints. One can use any data set taken from Table 9 depending on the desire 
for accuracy, and apply it to stars whose iron abundances are available.

\subsubsection{Calibration of $M_J$ in terms of Metallicity}
We estimated the $M_J$ absolute magnitude for the $(V-J)_0$ colours given in Table 10 
for the cluster sample in Table 4 by combining the $J_0$ apparent magnitude evaluated by 
Eq. (6) and the true distance modulus ($\mu_0$) of the cluster in question, i.e.
\begin{eqnarray}
M_J= J_0-\mu_0.                     
\end{eqnarray}
The absolute magnitudes versus $(V-J)_0$ colours are plotted in Fig. 7. We fitted the 
$M_J$ absolute magnitudes to the corresponding $[Fe/H]$ metallicity for the following 
$(V-J)_0$ colour indices and obtained the required calibrations just for the exhibition 
of the procedure: $(V-J)_0=1.00$, 1.15, 1.30, 1.50, and 1.70. The results are given in 
Table 11 and Fig. 8. The absolute magnitudes in all colour indices are fitted to a 
quadratic polynomial with (squared) correlation coefficients ($R^2$) between 0.9667 
and 1, where the first one corresponds to the absolute magnitudes for the colour 
index $(V-J)_0=1.15$ mag. 

We adopted the interval in $0.90<(V-J)_0\leq1.75$ mag, where at least three clusters 
are defined, for the application of the procedure and we evaluated $M_J$ absolute 
magnitudes for each colour. Then, we combined them with the corresponding $[Fe/H]$ 
metallicities and obtained the final calibrations. The general form of the equation 
for the calibrations is as follows:

\begin{eqnarray}
M_J = d_0 + d_1 X+ d_2X^2,                    
\end{eqnarray} 
where $X=[Fe/H]$. The absolute magnitudes estimated via Eq. (10) for 85 $(V-J)_o$ 
colour indices and the corresponding $d_i$ ($i=0$, 1, 2) coefficients are given 
in Table 12. 

\subsection{Application of the Procedure}
We have two absolute magnitude calibrations, $M_g$ and $M_J$, in terms of metallicity. 
Hence, we can apply each of them to a set of data.

\subsubsection{Application of the $M_g \times \lbrack Fe/H\rbrack$ Calibration} 
The procedure is applied to four clusters with different metal abundances (M53, M3, 
M71, and M35). The reason of preferring stellar clusters instead of individual field dwarfs 
is that clusters provide absolute magnitudes for comparison with the ones estimated 
by the method. The choice of test clusters as well as the calibrator ones is arbitrary.
Hence, we did not involve these clusters into the calibration set but 
we used them to confirm the robustness of the procedure. In the case of field stars, 
one needs their distances or trigonometric parallaxes in addition to their $g_0$ and 
$(g-r)_0$ data in order to evaluate their absolute magnitudes for their comparison 
with the ones estimated by the procedure in our work. However, such stars are rare 
in the literature. The parameters of stellar clusters and ($g$, $g-r$) photometric data  
are given in Table 13 and Table 14, respectively. The colour-apparent magnitude diagrams of the 
Galactic clusters used for the application of the procedure are shown in Fig. 9. 
The $g$ and the $g-r$ data of the clusters M53, M3, and M71 are taken from \cite{An08}, 
whereas those of M35 are transformed from the $V$ and $B-V$ data in \cite{vonHippel02} 
by the reduced transformation equations of \cite{Chonis08} given in the following:

\begin{eqnarray}
g=V+0.642\times(B-V)-0.135,\nonumber\\ 
g-r= 1.094\times(B-V)-0.248. 
\end{eqnarray} 
The reason of this choice is to obtain the $g$ and $g-r$ data for a relatively 
metal rich cluster with a large colour range. The metallicity of M35 is 
$\lbrack Fe/H\rbrack=-0.16$ dex, and it covers the colour range of $-0.30 
\leq(g-r)_0\leq 1.40$ mag. However, the application could be carried out 
for the colour interval $0.35\leq(g-r)_{0}\leq1.25$ mag, due to the restriction 
of our calibrations for the metallicity $\lbrack Fe/H\rbrack=-0.16$ dex (see Table 9). 

We evaluated the $M_g$ absolute magnitude by Eq. (8) for a set of $(g-r)_0$ 
colour indices where the clusters are defined. The results are given in Table 15. 
The columns refer to (1) $(g-r)_0$, colour index; (2) $(M_g)_{ev}$, the absolute 
magnitude estimated by the procedure; (3) $(M_g)_{cl}$, absolute magnitude for 
a cluster estimated  by its colour-magnitude diagram; and (4) $\Delta M$, 
absolute magnitude residuals. Also, the metallicity for each cluster is indicated 
near the name of the cluster. The differences between the absolute magnitudes 
estimated by the procedure presented in this study and those evaluated via colour-magnitude 
diagrams of the clusters (the residuals) lie between $-0.15$ and $+0.12$ mag. 
The mean and the standard deviation of the residuals are $<\Delta M_g>=-0.002$ and 
$\sigma=0.065$ mag, respectively. The distribution of the residuals is given in 
Table 16 and Fig. 10. 

The linear equation (Eq. 9) is valid for the colours $(g-r)_0 \geq 0.82$. Hence we 
could applied it to only cluster M35. We used the $c_1$ and $c_0$ coefficients in Table 9 
and evaluated the $(M_g)_{ev}$ absolute magnitudes for nine $(g-r)_0$ colours, i.e. 
0.85, 0.90, 0.95, 1.0, 1.05, 1.10, 1.15, 1.20, and 1.25 mag. Then we compared them 
with the corresponding ones, $(M_g)_{cl}$. The mean of the residuals and the standard 
deviation for this set of data are $<\Delta M>=-0.102$ and $\sigma=0.094$ mag, 
respectively. 

\subsubsection{Application of the $M_J \times \lbrack Fe/H\rbrack$ Calibration}
We applied the procedure to the cluster M5 and to a set of field stars with solar 
metallicity as explained in the following. The reason of choosing a cluster is that 
it provides absolute magnitudes for comparison with the ones estimated by the procedure. 
The colour excess, $E(B-V)=0.03$ mag and the metallicity $\lbrack Fe/H\rbrack=-1.40$ dex 
of the cluster are taken from \cite{Reid97}, while for the apparent distance modulus 
we adopted the one of \cite{Brasseur10}, $\mu=14.45$ mag. \cite{Sarajedini04} published 
the 2MASS magnitudes, $V$ magnitudes, metallicities and parallaxes of 54 field stars. 
Parallaxes provide absolute magnitudes for comparison with the ones estimated by our 
procedure, and this is the reason of choosing this sample of stars. The range of the 
metallicity is $-0.45\leq \lbrack Fe/H\rbrack \leq +0.35$ dex, i.e. the stars are of solar 
metallicity. Hence, their combination with (relatively) metal-poor stars in cluster 
M5 provides a good sample for the application of the procedure. Two stars, Hip 69301 
and Hip 84164, are omitted due to their large relative parallax errors, $\sigma_\pi/\pi=0.16$ 
and 0.09, respectively. Also, six stars which fall off the colour domain of our procedure, 
i.e. $(V-J)_0>1.75$ mag, could not be included into our programme. The $V-J$ colours and 
$J$ magnitudes of M5 cluster were de-reddened by the equations given in Section 3.1.2, 
while all the magnitudes in \cite{Sarajedini04} were assumed as unaffected from 
the interstellar extinction due to the proximity of the field stars to the Sun, $d<75$ pc. 
The results are given in Table 17 and Fig. 11. The data of the field stars in columns 
(1)-(6) are original, while those in (7), (8) and (9) are evaluated in this paper. 
The distance (d) was evaluated by using the corresponding parallax, and the absolute 
magnitude $M_J$ was calculated by the well known Pogson formula.
      
We evaluated the $M_J$ absolute magnitude by Eq. (11) for two sets of $(V-J)_0$ colour 
indices. The first set covers the domain of the cluster M5 (Table 18a), while the second 
one consists of $(V-J)_0$ colour indices of the field stars (Table 18b). The columns in 
Table 18a refer to (1) $(V-J)_0$ colour index, (2) $(M_J)_{cl}$, the absolute magnitude 
estimated by the combination of the colour-magnitude diagram and the true distance modulus 
of the cluster M5, (3) $(M_J)_{ev}$, the absolute magnitude estimated by the procedure, 
and (4) $\Delta$M, absolute magnitude residuals. For metallicity we adopted the one 
of cluster M5, i.e. $\lbrack Fe/H\rbrack=-1.40$ dex. The columns in Table 18b refer 
to (1) $(V-J)_0$ colour index of the field star, (2) $(M_J)_\pi$, the absolute magnitude 
estimated by the parallax of the field star, (3) $\lbrack Fe/H\rbrack$, the metallicity 
of the field star, (4) $(M_J)_{ev}$, the absolute magnitude estimated by the procedure, 
and (5) $\Delta M$, absolute magnitude residuals. The total residuals lie between $-0.29$ 
and $+0.35$ mag. The extreme residuals $\Delta M=-1.03$ and +0.56 mag which are 
marked in boldface in Table 18b correspond to stars Hip 84164 and Hip 69301 whose relative 
parallax errors are high, as mentioned above. Hence, they are not considered in the 
calculations of the mean residual and standard deviation. However, the range of 87 per 
cent of them is rather shorter, i.e. $-0.20<\Delta M \leq+0.20$ mag. The mean and the 
standard deviation of all residuals, except the two extreme ones, are $<\Delta M_J>=0.05$ 
and $\sigma=0.13$ mag. The distribution of the residuals is also given in Table 19 and 
Fig. 12. We state that the cluster fiducials provide more accurate absolute magnitudes 
than the ones for the field stars. Actually, the mean and the standard deviation of the 
residuals for the fiducials are $<\Delta M_J>=0.04$ and $\sigma=0.05$ mag, whereas those 
for the field stars are $<\Delta M_J>=0.06$, $\sigma=0.16$ mag, respectively. 

\section{Summary and Discussion}
We presented two absolute magnitude calibrations for dwarfs based on the colour 
magnitude diagrams of Galactic clusters with different metallicities. For the 
calibration of $M_g$, we used the clusters M92, M13, M5, NGC 2420, M67, and 
NGC 6791. We combined the calibrations between $g_0$ and $(g-r)_0$ for each 
cluster with their true distance modulus and evaluated a set of absolute 
magnitudes for the $(g-r)_0$ range of each cluster. Then, we fitted the $M_g$ 
absolute magnitudes in terms of the iron metallicity, $[Fe/H]$, by a quadratic 
polynomial for a given $(g-r)_0$ colour index. Our absolute magnitude 
calibrations cover the range $0.25\leq(g-r)_0\leq1.25$ mag.	

We applied the procedure to another set of Galactic cluster, i.e. M53, M3, 
M71, and M35 and compared the absolute magnitudes estimated by this procedure 
with those evaluated via a combination of the fiducial $g_0$, $(g-r)_0$ 
sequence and the true distance modulus for each cluster. The residuals 
lie between $-0.15$ and $+0.12$ mag, and their mean and standard deviations 
are $<\Delta M>= -0.002$ and $\sigma=0.065$ mag. The range of the residuals 
estimated for the red giants with SDSS was $-0.28 \leq \Delta M \leq +0.43$ 
mag \citep{Karaali13a}, larger than the one for the dwarfs estimated in this 
study. Also, their mean and standard deviation were larger, i.e. $<\Delta M>=0.169$ 
and $\sigma= 0.140$. The difference between two sets of residuals originates 
mainly from the different trends of the colour-magnitude diagrams of 
red giants and dwarfs, i.e. the colour-magnitude diagram of red giants 
for a given cluster is steeper than the one for dwarfs of the same cluster, 
and any error in $(g-r)_0$ results in larger colour errors within the 
steeper diagram. 
 
The range of the $(B-V)_0$ colour in \cite{Karaali03} who estimated absolute 
magnitudes for dwarfs by absolute magnitude offsets is $0.3\leq(B-V)_0\leq1.1$ 
mag which corresponds to $0.1\leq(g-r)_0\leq0.9$ mag, shorter than the one 
cited in our study, i.e. $0.25\leq(g-r)_0\leq1.25$ mag. That is, there is 
an improvement on our procedure with respect to the one of \cite{Karaali03}. 

For the calibration of $M_J$, we used the clusters M92, M13, 47 Tuc, NGC 2158 
and NGC 6791. We combined the calibrations between $J_0$ and $(V-J)_0$ 
for each cluster with their true distance modulus and evaluated a set of 
absolute magnitudes for the $(V-J)_0$ range of each cluster. Then, 
we calibrated the MJ absolute magnitudes in terms of the iron metallicity, 
$[Fe/H]$, by a quadratic polynomial for all cluster. Our absolute 
magnitude calibrations cover the range $0.90<(V-J)_0\leq1.75$ mag. 

This is our fourth paper devoted to the calibration of the absolute 
magnitudes based on the colour-magnitude diagrams of Galactic clusters, 
and it is the first time that we used a set of field stars, additional 
to the Galactic cluster M5 ($[Fe/H]=-1.40$ dex), for the application 
of the $M_J \times[Fe/H]$ calibration. We tried to apply the procedure 
to a metal-poor cluster, i.e. M15 whose metallicity is close to the 
one of M92. But we noticed that the residuals were decreasing systematically 
from $\Delta M=0.51$ to $\Delta M=-0.26$ mag in the colour range 
$0.95\leq(V-J)_0\leq 1.40$ mag. Comparison of the absolute magnitude-colour 
diagrams of the clusters M92 and M15 (Fig. 11c) revealed that the reason for 
the systematic variation of the residuals is due to the fiducial sequence 
of the cluster M15. Actually, although the absolute magnitudes evaluated for 
M92 and M15 for the colour $(V-J)_0=0.95$ mag are close to each other, 
the absolute magnitudes of M15 deviate from the ones of M92 up to $\Delta M=0.51$ 
mag at redder colours. Hence, the systematic residuals in question do not 
originate from our procedure but they are due to the fiducial sequence of 
the cluster M15. The fiducial sequences of the clusters M92 and M15 
are taken from the same source, i.e. \cite{Brasseur10}. The metallicities 
for both clusters are given as $[Fe/H]=-2.4$ dex in the same paper. Additionally, 
they have the same age, a bit less than 13 Gyr \citep{Salaris97}. Hence, 
the problem related to the discrepancy between the two fiducial 
sequences could not be solved.

We compared the absolute magnitudes estimated by the procedure explained 
in our paper with those evaluated either by combination of the $J_0$, $(V-J)_0$ 
fiducial sequence of the cluster M5 and its true distance modulus or by the 
$J_0$ apparent magnitude and the distance of the field star in question. 
The total residuals lie between $-0.29$ and $+0.39$ mag. However, the range 
of 87 per cent of them is rather shorter, i.e. $-0.20 < \Delta M \leq +0.20$ mag. 
The mean and the standard deviation of all residuals are $<\Delta M>=0.05$ and 
$\sigma=0.13$ mag. However, we state that the cluster fiducials provide more 
accurate absolute magnitudes than the ones for the field stars. Thus, the 
absolute magnitude $M_J$ of a dwarf would be determined with an accuracy 
of $\Delta M < 0.2$ mag. The mean and standard deviation of the residuals 
estimated with the same photometry $(V, J)$ for red giants were $<\Delta M>=0.137$ 
and $\sigma=0.080$ mag \citep{Karaali13b} which indicates that the absolute 
magnitudes of dwarfs would be determined with better accuracy. The reason of 
this improvement is the same as stated for the absolute magnitude $M_g$. 
The calibration of $M_g$ and $M_J$ in terms of $[Fe/H]$ is carried out 
in steps of 0.01 mag. A small step is necessary to isolate an observational 
error on $g-r$ and $V-J$ plus an error due to reddening. The origin of 
the mentioned errors shows the trend of the dwarf sequence. However, we note 
that the dwarf sequence is not steep. Hence, contrary to the red giant branch 
sequence \cite{Karaali13a}, an error in $(g-r)$ or $(V-J)$ does not imply 
a large change in the absolute magnitude. 

The $c_i$ and $d_i$ ($i=$ 0, 1, 2) coefficients in Tables 9 and 12, 
respectively, are colour dependent. Additionally, it seems that there is a 
smooth relation between a given $c_i$ or $d_i$ coefficient and the $(g-r)$ 
and $(V-J)$ colour. We confirmed this argument by plotting each $c_i$ 
coefficient versus $(g-r)_0$ colour. Actually, each panel in Fig. 13 consists 
of a set of smooth relations with breaks at $(g-r)_0$ colours where the number 
of clusters employed in the $\lbrack Fe/H\rbrack$ calibrations changes. It is 
interesting, the mentioned breaks in the panel $c_0$ versus $(g-r)_0$ are so 
small that one can assume a continuous relation for the whole colour interval, 
i.e. 0.25 $\leq(g-r)_0\leq$ 1.25 mag. The mentioned relation originates from 
the fact that $M_g=c_0$ for $\lbrack Fe/H\rbrack=0$ dex.

The procedure for absolute magnitude calibration for dwarfs in \cite{Karaali03} is based on 
the absolute magnitude offset which is defined as a function of ultraviolet excess. However, 
ultraviolet excess can not be provided easily with SDSS, especially for late-type stars, 
and this parameter has not been defined in 2MASS system. Whereas, the procedure in this 
study is metallicity dependent, and metallicity values come from detailed analysis of 
individual stars using high – resolution spectroscopy in large surveys 
which are in operation, such as  RAVE, SEGUE, {\em GAIA}-ESO, and APOGEE or which 
will be started in the near future, such as GALAH. 

SDSS and 2MASS are two of the widely used photometric systems in the Galactic researches. 
Also, our procedure covers dwarfs with a large metallicity and age range. The metallicity 
range is $-2.15\leq \lbrack Fe/H\rbrack \leq0.37$ dex in both systems, while the age ranges 
are defined by the clusters used for the calibrations in SDSS and 2MASS, i.e. $4\leq t \leq13.2$ 
and $2\leq t \leq 13.2$ Gyr, respectively, where 2, 4 and 13.2 Gyr are the ages of the 
clusters NGC 2158, M67 and M92. The age of the cluster NGC 2158 is taken from 
\cite{Carraro02}, while those for the clusters M67 and M92 were estimated in 
\cite{Karaali12}. Additionally, the mean residuals and standard deviations 
in the application of two absolute magnitude calibrations, $M_g$ and $M_J$, 
are small. Hence, the procedure presented in this paper can be applied 
with a good accuracy to large samples of dwarf stars in Galactic-structure Projects. 

We stated in our previous papers \citep{Karaali12} that our absolute magnitude 
calibrations are age-restricted, i.e. they cover the stars with ages in the range 
defined by the youngest and oldest clusters used in the calibration. We applied the 
calibration $M_g \times \lbrack Fe/H\rbrack$ to two clusters, NGC 3680 and NGC 2158, 
younger than 4 Gyr to test the age range in the SDSS system. We used the $V$ and 
$B-V$ data of the cluster NGC 3680 in \cite{Kozhurina97} and reduced  them to the 
$g_0$ magnitude and $(g-r)_0$ colour by means of the equations of \cite{Chonis08} 
given in Section 3.2.1, after necessary corrections for interstellar extinction. 
The colours and magnitudes of the cluster NGC 2158 taken from \cite{Smolinski11} 
are already in the SDSS system. The $M_g \times \lbrack Fe/H\rbrack$ calibration 
could be applied to 9 stars in NGC 3680 and 10 fiducials in NGC 2158, and the 
absolute magnitude residuals for each set of data were estimated. The mean residuals 
are very similar and rather high, i.e. $<\Delta M> = 1.25$ mag for 
9 stars and $<\Delta M> = 1.20$ mag. for 10 fiducials, which shows that the calibration 
in question is limited with the age of the youngest cluster used for the calibration, 
i.e. 4 Gyr. The metallicities of the clusters NGC 3680 and NGC 2158 are close to 
the one of M67 whose age is 4 Gyr, while the ages of the first two clusters are 
different than the one of M67. A similar test can be carried out for the 
$M_J \times \lbrack Fe/H\rbrack$ calibration. Hence, we can argue that our 
absolute magnitude calibrations are better defined in terms of ranges in age. 
The colour excesses, distance moduli, metallicities and ages of the clusters 
are presented in Table 20. However, the magnitudes and colours used in the evaluation 
of absolute magnitude residuals are not given in the paper because of space constraints.                

\section*{Acknowledgments} 
We would like to thank the anonymous referee who provided valuable comments and  
for improving the manuscript. This research has made use of NASA(National Aeronautics 
and Space Administration)'s Astrophysics Data System and the SIMBAD Astronomical 
Database, operated at CDS, Strasbourg, France.

\begin{table*}
\setlength{\tabcolsep}{4pt}
\center
\caption{Data for the cluster used in the calibration with SDSS.}
\begin{tabular}{lcccc}
    \hline
    Cluster    &$E(B-V)$&$\mu_0$& $\lbrack Fe/H\rbrack$&Ref \\
    \hline
		& (mag) & (mag) & (dex) &  \\
\hline
    M92   	& 0.025 & 14.72 & -2.15 & 1 \\
    M13   	& 0.020 & 14.38 & -1.41 & 1 \\
    M5    	& 0.040 & 14.37 & -1.26 & 2 \\
    NGC 2420 	& 0.040 & 12.00 & -0.37 & 2 \\
    M67   	& 0.038 &  9.53 & -0.04 & 1 \\
    NGC 6791 	& 0.150 & 13.10 & +0.37 & 3 \\
    \hline
    \end{tabular}\\
(1) \cite{Gratton97}, (2) \cite{An08}, (3) \cite{Brasseur10}\\ 
\end{table*}

\begin{table*}[h]
\setlength{\tabcolsep}{4pt}
\center
\tiny{
\caption{$g$, $r$ magnitudes, $g-r$ and de-reddened $(g-r)_0$ colours and $g_0$ magnitudes for the clusters M92, M13, M5, NGC 2420, M67, and NGC 6791 used in the calibration with SDSS.}
\begin{tabular}{ccccc|ccccc}
    \hline
     $r$ & $g-r$ & $g$ & $(g-r)_0$ & $g_0$ & $r$ & $g-r$ & $g$ & $(g-r)_0$ & $g_0$ \\
    \hline
    \multicolumn{5}{c}{M92}               & \multicolumn{5}{|c}{M5} \\
    \hline
    18.750 & 0.225 & 18.975 & 0.199 & 18.882 & 19.850 & 0.411 & 20.261 & 0.369 & 20.112 \\
    19.050 & 0.240 & 19.290 & 0.214 & 19.197 & 20.150 & 0.463 & 20.613 & 0.421 & 20.464 \\
    19.350 & 0.263 & 19.613 & 0.237 & 19.520 & 20.450 & 0.522 & 20.972 & 0.480 & 20.823 \\
    19.650 & 0.291 & 19.941 & 0.265 & 19.848 & 20.750 & 0.588 & 21.338 & 0.546 & 21.189 \\
    19.950 & 0.326 & 20.276 & 0.300 & 20.183 & 21.050 & 0.665 & 21.715 & 0.623 & 21.566 \\
    20.250 & 0.369 & 20.619 & 0.343 & 20.526 & 21.350 & 0.754 & 22.104 & 0.712 & 21.955 \\
\hline
\end{tabular}
}
\end{table*}

\begin{table*}[h]
\setlength{\tabcolsep}{3pt}
\center
\caption{The values of the coefficients $a_i$ ($i=$0,...,5) in Eq. (1).}
\begin{tabular}{rrrrrrr}
    \hline
    Cluster & M92   & M13   & M5    & NGC 2420 & M67   & NGC 6791 \\
    \hline
    $(g-r)_0$ interval & [0.20, 0.46] & [0.25, 0.78] & [0.25, 0.81] & [0.19, 1.40] & [0.33, 1.40] & [0.51, 1.27] \\
\hline
    $a_5$   &   7289.856 &  224.151 &  259.663 &   8.460 &  13.044 &   53.081 \\
    $a_4$   & -12324.726 & -632.530 & -732.851 & -32.353 & -52.757 & -250.015 \\
    $a_3$   &   8269.137 &  708.233 &  813.566 &  50.359 &  86.340 &  468.511 \\
    $a_2$   &  -2768.241 & -396.039 & -447.797 & -41.975 & -73.637 & -437.308 \\
    $a_1$   &    472.894 &  116.688 &  128.576 &  23.339 &  37.436 &  207.802 \\
    $a_0$   &    -13.697 &    5.748 &    4.585 &  11.978 &   6.911 &  -21.621 \\
\hline
\end{tabular}
\end{table*}

\begin{table*}
\setlength{\tabcolsep}{3pt}
\center
\caption{Data for the clusters used in the calibration with 2MASS.}  
\begin{tabular}{ccccc}
\hline
   Cluster &     $E(B-V)$ &   $\mu_0$&     $[Fe/H]$ & References \\
           &      (mag) &      (mag) &      (dex) &            \\
\hline
       M92 &      0.025 &      14.72 &      -2.15 &          1 \\
       M13 &      0.020 &      14.38 &      -1.41 &          1 \\
    47 Tuc &      0.220 &      15.40 &      -1.27 &       2, 3 \\
  NGC 2158 &      0.440 &      13.38 &      -0.25 &       4, 5 \\
  NGC 6791 &      0.150 &      13.10 &       0.37 &          6 \\
\hline
\end{tabular}
\\
(1) \citet{Gratton97}, (2) \citet{Harris96}, (3) \citet{Alcaino90}, (4) \citet{Smolinski11}, (5) \citet{Arp62}, (6) \citet{Brasseur10}  
\end{table*}

\begin{table*}
\setlength{\tabcolsep}{3pt}
{\tiny
\center
\caption{Colours and magnitudes for the clusters used in the calibration with 2MASS.} 
\begin{tabular}{ccccccccccc}
\hline
       M92 &            &            &            &            &        M13 &            &            &            &            &            \\
$V$ & $V-J$ & $J$ &$(V-J)_0$ &  $J_0$ & $V$ & $V-J$ &   $J$ & $(V-J)_0$ & $J_0$ & \\
\hline
21.994 &      1.883 &     20.111 &      1.827 &     20.089 &     22.039 &      2.011 &    20.028 &      1.966 &     20.011 &    $-$        \\
21.836 &      1.752 &     20.084 &      1.696 &     20.062 &     21.884 &      1.939 &     19.945 &      1.894 &     19.928 &   $-$         \\
21.688 &      1.684 &     20.004 &      1.628 &     19.982 &     21.740 &      1.883 &     19.857 &      1.838 &     19.840 &   $-$         \\
21.550 &      1.621 &     19.929 &      1.565 &     19.907 &     21.606 &      1.832 &     19.774 &      1.787 &     19.757 &   $-$         \\
21.420 &      1.570 &     19.850 &      1.514 &     19.828 &     21.460 &      1.766 &     19.694 &      1.721 &     19.677 &   $-$         \\
21.187 &      1.474 &     19.713 &      1.418 &     19.691 &     21.306 &      1.709 &     19.597 &      1.664 &     19.580 &   $-$         \\
\hline
\end{tabular}
}  
\end{table*}

\begin{table*}
\setlength{\tabcolsep}{3pt}
\center
\caption{Numerical values of the coefficients $b_i$ ($i$=0, 1, 2, 3, 4, 5).} 
\begin{tabular}{crrrrr}
\hline
 Cluster &  M92 &   M13 & 47 Tuc &   NGC 2158 &   NGC 6791 \\
$(V-J)_0$ interval & [0.90, 1.83] & [0.96, 1.96] & [0.90, 1.37] & [0.57, 1.31] & [1.19, 1.72] \\
\hline
 $b_5$ &     13.498 &     11.536 &     152.580 &       --- &     74.861 \\
 $b_4$ &    -97.365 &    -88.498 &    -957.780 &       --- &   -583.990 \\
 $b_3$ &    278.950 &    269.890 &    2395.100 &    -1.791 &   1819.900 \\
 $b_2$ &   -399.070 &   -409.920 &   -2982.400 &     3.395 &  -2832.700 \\
 $b_1$ &    288.640 &    312.580 &    1852.200 &     2.183 &   2204.800 \\
 $b_0$ &    -66.255 &    -77.700 &    -440.800 &    12.745 &   -670.490 \\
\hline
\end{tabular}  
\end{table*}

\begin{table*}
\setlength{\tabcolsep}{5pt}
\center
\caption{$M_g$ absolute magnitudes estimated for a set of $(g-r)_0$ colours for six clusters used in the calibration with SDSS.}
\begin{tabular}{ccccccc}
\hline
		& M92   & M13   & M5  & NGC 2420 & M67   & NGC 6791 \\
      $(g-r)_0$ & $M_g$ & $M_g$ &$M_g$&  $M_g$ & $M_g$   & $M_g$    \\
   \hline
    0.20  & 4.199 & $-$   & $-$   & 3.321 & $-$  & $-$ \\
    0.25  & 4.972 & 4.603 & 4.475 & 3.858 & $-$  & $-$ \\
    0.30  & 5.460 & 5.275 & 5.148 & 4.320 & $-$  & $-$ \\
    0.35  & 5.864 & 5.746 & 5.611 & 4.723 & 4.442 & $-$ \\
    0.40  & 6.178 & 6.107 & 5.965 & 5.079 & 4.883 & $-$ \\
    0.45  & 6.462 & 6.417 & 6.273 & 5.399 & 5.261 & $-$ \\
    0.50  & $-$   & 6.704 & 6.563 & 5.691 & 5.593 & $-$ \\
    0.55  & $-$   & 6.978 & 6.840 & 5.961 & 5.890 & 5.024 \\
    0.60  & $-$   & 7.240 & 7.100 & 6.213 & 6.160 & 5.451 \\
    0.65  & $-$   & 7.485 & 7.333 & 6.450 & 6.410 & 5.779 \\
    0.70  & $-$   & 7.717 & 7.537 & 6.675 & 6.644 & 6.048 \\
    0.75  & $-$   & 7.953 & 7.728 & 6.887 & 6.865 & 6.283 \\
    0.80  & $-$   & $-$   & 7.945 & 7.089 & 7.073 & 6.503 \\
    0.85  & $-$   & $-$   & $-$   & 7.281 & 7.271 & 6.717 \\
    0.90  & $-$   & $-$   & $-$   & 7.464 & 7.458 & 6.927 \\
    0.95  & $-$   & $-$   & $-$   & 7.639 & 7.636 & 7.136 \\
    1.00  & $-$   & $-$   & $-$   & 7.808 & 7.807 & 7.340 \\
    1.05  & $-$   & $-$   & $-$   & 7.976 & 7.975 & 7.539 \\
    1.10  & $-$   & $-$   & $-$   & 8.146 & 8.145 & 7.734 \\
    1.15  & $-$   & $-$   & $-$   & 8.327 & 8.324 & 7.930 \\
    1.20  & $-$   & $-$   & $-$   & 8.526 & 8.523 & 8.139 \\
    1.25  & $-$   & $-$   & $-$   & 8.755 & 8.757 & 8.381 \\
    1.30  & $-$   & $-$   & $-$   & 9.028 & 9.043 & $-$ \\
    1.35  & $-$   & $-$   & $-$   & 9.363 & 9.402 & $-$ \\
\hline
\end{tabular}
\end{table*}

\begin{table*}
\setlength{\tabcolsep}{5pt}
\center
\caption{$M_g$ absolute magnitudes and $\lbrack Fe/H\rbrack$ metallicities 
for six $(g-r)_0$ intervals.}
\begin{tabular}{ccc|ccc}
\hline
$(g-r)_0$&$[Fe/H]$&$M_g$&$(g-r)_0$ & $[Fe/H]$ & $M_g$ \\
\hline
    (mag) & (dex) & (mag) & (mag) & (dex) & (mag) \\
\hline
     0.25 & -2.15 & 4.972 &  0.80 & -1.26 & 7.945 \\
          & -1.41 & 4.603 &       & -0.37 & 7.089 \\
          & -1.26 & 4.475 &       & -0.04 & 7.073 \\
          & -0.37 & 3.858 &       &  0.37 & 6.503 \\
\hline
     0.40 & -2.15 & 6.178 &  0.95 & -0.37 & 7.639 \\
          & -1.41 & 6.107 &       & -0.04 & 7.636 \\
          & -1.26 & 5.965 &       &  0.37 & 7.136 \\
\cline{4-6}
          & -0.37 & 5.079 &  1.20 & -0.37 & 8.526 \\
          & -0.04 & 4.883 &       & -0.04 & 8.523 \\
\cline{1-3}
     0.65 & -1.41 & 7.485 &       & 0.37  & 8.139 \\
          & -1.26 & 7.333 & $-$   & $-$   & $-$ \\
          & -0.37 & 6.450 & $-$   & $-$   & $-$ \\
          & -0.04 & 6.410 & $-$   & $-$   & $-$ \\
          &  0.37 & 5.779 & $-$   & $-$   & $-$ \\
\hline
\end{tabular}
\end{table*}

\begin{table*}
\setlength{\tabcolsep}{.5pt}
\center
\tiny{
\caption{Absolute magnitudes estimated for six Galactic clusters and the numerical values of $c_i$ ($i=$ 0, 1, 2) coefficients in Eqs. (8) and (9).}
\begin{tabular}{cccccccccccc|cccccccccccc}
\hline
& M92   & M13   & M5    & NGC 2420 & M67   & NGC 6791 &       &       &       &       &       &       & M92   & M13   & M5    & NGC 2420 & M67   & NGC 6791  & (cont.)  &   &   & & \\
 \hline
$(g-r)_0$ &  $M_g$ & $M_g$  &  $M_g$   & $M_g$ & $M_g$   & $M_g$  & $c_2$ &  $c_1$ &  $c_0$ &  $R^{2}$ &  $[Fe/H]$ & $(g-r)_0$ &  $M_g$ & $M_g$  &  $M_g$   & $M_g$ & $M_g$   & $M_g$  & $c_2$ &  $c_1$ &  $c_0$ &  $R^{2}$ &  $[Fe/H]$  \\
&   &   &     &  &    &  &       &       &       &       &  interval     &  &   &     &  &    &  &       &       &       &       &       & interval  \\
\hline
0.25& 4.972 & 4.603 & 4.475 & 3.858 & $-$ & $-$& -0.0977 & -0.8738 & 3.5469 & 0.9990 & [-2.15,-0.37] &0.76& $-$  & 8.004 & 7.767 & 6.929 & 6.908 & 6.328 & 0.0094 & -0.8681 & 6.7093 & 0.9754 & [-1.41, 0.37] \\
0.26& 5.081 & 4.759 & 4.633 & 3.956 & $-$& $-$& -0.1663 & -1.0527 & 3.5880 & 0.9991 & [-2.15,-0.37]&0.77& $-$  & 8.057 & 7.808 & 6.970 & 6.950 & 6.372 & 0.0182 & -0.8609 & 6.7507 & 0.9749 & [-1.41, 0.37] \\
0.27& 5.182 & 4.904 & 4.778 & 4.051 & $-$& $-$& -0.2255 & -1.2054 & 3.6346 & 0.9991 & [-2.15,-0.37] & 0.78  & $-$  & 8.113 & 7.851 & 7.010 & 6.991 & 6.416 & 0.0285 & -0.8538 & 6.7915 & 0.9744 & [-1.41, 0.37] \\
0.28& 5.278 & 5.037 & 4.912 & 4.143 & $-$& $-$& -0.2754 & -1.3335 & 3.6863 & 0.9991 & [-2.15,-0.37] & 0,79  & $-$  & $-$  & 7.896 & 7.050 & 7.033 & 6.460 & -0.0492 & -0.8850 & 6.8449 & 0.9625 & [-1.26, 0.37] \\
0.29& 5.371 & 5.161 & 5.035 & 4.233 & $-$& $-$& -0.3166 & -1.4388 & 3.7426 & 0.9991 & [-2.15,-0.37] & 0.80  & $-$  & $-$  & 7.945 & 7.089 & 7.073 & 6.503 & -0.0403 & -0.8803 & 6.8854 & 0.9625 & [-1.26, 0.37] \\
\hline
\end{tabular}
}
\end{table*}

\begin{table*}
\setlength{\tabcolsep}{3pt}
\center
\caption{$M_J$ absolute magnitudes estimated for a set of $(V-J)_0$ colours for five clusters used for the calibration with 2MASS. The absolute magnitudes with boldface are not considered in the calibration (see text).} 
\begin{tabular}{cccccc}
\hline
Cluster   &   M92 &     M13 &     47 Tuc &   NGC 2158 &   NGC 6791 \\
$(V-J)_0$ & $M_J$ &   $M_J$ &      $M_J$ &      $M_J$ &      $M_J$ \\
    (mag) & (mag) &   (mag) &      (mag) &      (mag) &      (mag) \\
\hline
0.90 &       3.00 &        --- &       2.76 &       2.77 &       ---  \\
0.95 &       3.38 &       3.16 &       3.22 &       2.97 &       ---  \\
1.00 &       3.68 &       3.51 &       3.50 &       3.15 &       ---  \\
1.05 &       3.92 &       3.78 &       3.69 &       3.33 &       ---  \\
1.10 &       4.12 &       3.99 &       3.84 &       3.49 &       ---  \\
1.15 &       4.30 &       4.16 &       3.98 &       3.64 &       ---  \\
1.20 &       4.45 &       4.30 &       4.13 &       3.78 &       3.19 \\
1.25 &       4.59 &       4.42 &       4.29 & {\bf 3.90} &       3.51 \\
1.30 &       4.71 &       4.53 &       4.44 & {\bf 4.00} &       3.73 \\
1.35 &       4.83 &       4.64 &       4.59 & {\bf 4.09} &       3.88 \\
1.40 &       4.94 &       4.74 &        --- & {\bf 4.16} &       4.01 \\
1.45 &       5.04 &       4.83 &        --- &       ---  &       4.12 \\
1.50 &       5.12 &       4.93 &        --- &       ---  &       4.22 \\
1.55 &       5.20 &       5.02 &        --- &       ---  &       4.33 \\
1.60 &       5.25 &       5.11 &        --- &       ---  &       4.43 \\
1.65 &       5.30 &       5.18 &        --- &       ---  &       4.52 \\
1.70 &       5.33 &       5.26 &        --- &       ---  &       4.61 \\
1.75 &       5.36 &       5.32 &        --- &       ---  &       4.71 \\
1.80 & {\bf 5.38} &  {\bf 5.39}&        --- &       ---  &       ---  \\
1.85 &       ---  &  {\bf 5.45}&        --- &       ---  &       ---  \\
1.90 &       ---  &  {\bf 5.51}&        --- &       ---  &       ---  \\
1.95 &       ---  &  {\bf 5.60}&        --- &       ---  &       ---  \\
\hline
\end{tabular}  
\end{table*}

\begin{table*}
\setlength{\tabcolsep}{3pt}
\center
\caption{$M_J$ absolute magnitudes and $[Fe/H]$ metallicities for five $(V-J)_0$ intervals.} 
\begin{tabular}{ccc}
\hline
  $(V-J)_0$ &  $[Fe/H]$ &     $M_J$ \\
    (mag) &      (dex) &      (mag) \\
\hline
1.00 &      -2.15 &       3.68 \\
     &      -1.41 &       3.51 \\
     &      -1.27 &       3.50 \\
     &      -0.25 &       3.15 \\
1.15 &      -2.15 &       4.30 \\
     &      -1.41 &       4.16 \\
     &      -1.27 &       3.98 \\
     &      -0.25 &       3.64 \\
1.30 &      -2.15 &       4.71 \\
     &      -1.41 &       4.53 \\
     &      -1.27 &       4.44 \\
     &      -0.25 &       4.00 \\
     &       0.37 &       3.73 \\
1.50 &      -2.15 &       5.12 \\
     &      -1.41 &       4.93 \\
     &       0.37 &       4.22 \\
1.70 &      -2.15 &       5.33 \\
     &      -1.41 &       5.26 \\
     &       0.37 &       4.61 \\
\hline
\end{tabular}  
\end{table*}

\begin{table*}
\setlength{\tabcolsep}{5pt}
\center
{\tiny
\caption{$M_J$ absolute magnitudes estimated for five Galactic clusters and the numerical values of $d_i$ ($i$=0, 1, 2) coefficients in Eq. 11.} 
\begin{tabular}{ccccccccccc}
\hline
 Cluster &        M92 &        M13 &     47 Tuc &   NGC 2158 &   NGC 6791 &         &         &         &      &  $[Fe/H]$ \\
 $(V-J)_0$ &   $M_J$ &     $M_J$ &       $M_J$ &     $M_J$ &     $M_J$ &   $d_2$   &   $d_1$ &   $d_0$    & $R^2$&   interval (dex)  \\
\hline
 0.91 &       3.08 &       ---  &       2.87 &       2.81 &     ---    &     0.0946 &     0.0856 &      2.8282 &      1.000 & [-2.15,-0.25] \\
 0.92 &       3.16 &       ---  &       2.97 &       2.85 &     ---    &     0.0507 &    -0.0411 &      2.8382 &      1.000 & [-2.15,-0.25] \\
 0.93 &       3.24 &       ---  &       3.06 &       2.89 &     ---    &     0.0150 &    -0.1461 &      2.8529 &      1.000 & [-2.15,-0.25] \\
 0.94 &       3.31 &       ---  &       3.14 &       2.93 &     ---    &    -0.0133 &    -0.2318 &      2.8715 &      1.000 & [-2.15,-0.25] \\
 0.95 &       3.38 &       3.16 &       3.22 &       2.97 &     ---    &     0.0122 &    -0.1831 &      2.9226 &      0.953 & [-2.15,-0.25] \\
 0.96 &       3.44 &       3.24 &       3.29 &       3.00 &     ---    &    -0.0086 &    -0.2482 &      2.9450 &      0.967 & [-2.15,-0.25] \\
\hline
\end{tabular}
}  
\end{table*}  

\begin{table*}
\setlength{\tabcolsep}{2pt}
\center
\caption{Data for the clusters used for the application of the procedure with SDSS.}
\begin{tabular}{ccccc}
\hline
    Cluster & $E(B-V)$ & $\mu_0$ & $[Fe/H]$ & Ref \\
              & (mag) & (mag) & (dex) &  \\
\hline
    M53   & 0.021 & 16.21 & -2.06 & 1,2 \\
    M3    & 0.01  & 14.83 & -1.46 & 3,4,5 \\
    M71   & 0.32  & 13.02 & -0.73 & 3 \\
    M35   & 0.19  & 9.62  & -0.16 & 6 \\
    \hline
    \end{tabular}\\
(1) \cite{Smolinski11}, (2) \cite{Ferro11}, (3) \cite{An08}, (4) \cite{Saad01}, 
(5) \cite{Di04}, (6) \cite{Sarajedini03}\\ 
\end{table*}

\begin{table*}
\setlength{\tabcolsep}{2pt}
\center
\caption{Fiducial dwarf sequence for the galactic clusters used in the application of the procedure with SDSS.}
\begin{tabular}{cccc|cccccc}
\hline
    $g-r$ & $g$     & $(g-r)_0$ & $g_0$    & $B-V$   & $V$     & $(B-V)_0$ & $V_0$ & $(g-r)_0$ & $g_0$ \\
\hline
   \multicolumn{4}{c|}{M53}       & \multicolumn{6}{|c}{M35} \\
   \hline
    0.263 & 21.113 & 0.241 & 21.035 & 0.70  & 14.42 & 0.51  & 13.83 & 0.31  & 14.02 \\
    0.291 & 21.441 & 0.269 & 21.363 & 0.80  & 15.06 & 0.61  & 14.47 & 0.42  & 14.73 \\
    0.324 & 21.774 & 0.302 & 21.696 & 0.90  & 15.61 & 0.71  & 15.02 & 0.53  & 15.34 \\
    0.362 & 22.112 & 0.340 & 22.034 & 1.00  & 16.00 & 0.81  & 15.41 & 0.64  & 15.80 \\
   \cline{1-4} 
   \multicolumn{4}{c|}{M71}         & 1.10  & 16.52 & 0.91  & 15.93 & 0.75  & 16.38 \\
   \cline{1-4}
    0.658 & 19.008 & 0.320 & 17.819 &  1.19 & 16.85 & 1.00  & 16.26 & 0.85  & 16.77 \\
\hline
\end{tabular}
\end{table*}

\begin{table*}
\setlength{\tabcolsep}{2pt}
\center
\caption{Absolute magnitudes ($(M_g)_{ev}$) and residuals ($\Delta M$) estimated by the procedure explained in our work. $(M_g)_{cl}$ denotes the absolute magnitude evaluated by means of colour - magnitude diagram of the cluster.}
\begin{tabular}{cccc|cccc|cccc}
\hline
    1     & 2     & 3     & 4     & 1     & 2     & 3     & 4     & 1     & 2     & 3     & 4 \\
     \hline
 $(g-r)_0$&$(M_g)_{ev}$&$(M_g)_{cl}$&$\Delta M$&$(g-r)_0$&$(M_g)_{ev}$&$(M_g)_{cl}$&$\Delta M$&$(g-r)_0$&$(M_g)_{ev}$&$(M_g)_{cl}$&$\Delta M$\\
\hline
    \multicolumn{4}{c}{M53 ($[Fe/H]=-2.06$ dex)} & \multicolumn{4}{|c}{M3 ($[Fe/H]=-1.46$ dex)} & \multicolumn{4}{|c}{M71 ($[Fe/H]=-0.73$ dex)} \\
    \hline
    0.25  & 4.932 & 4.936 & -0.004 & 0.35  & 5.718 & 5.671 &  0.047 & 0.32  & 4.920 & 4.806 &  0.114 \\
    0.26  & 5.051 & 5.051 & -0.001 & 0.36  & 5.792 & 5.753 &  0.039 & 0.33  & 5.008 & 4.936 &  0.072 \\
    0.27  & 5.161 & 5.163 & -0.002 & 0.37  & 5.863 & 5.833 &  0.031 & 0.34  & 5.092 & 5.059 &  0.033 \\
    0.28  & 5.265 & 5.270 & -0.005 & 0.38  & 5.931 & 5.910 &  0.021 & 0.35  & 5.199 & 5.176 &  0.023 \\
    0.29  & 5.363 & 5.373 & -0.010 & 0.39  & 5.997 & 5.986 &  0.011 & 0.36  & 5.279 & 5.286 & -0.007 \\
\hline
\end{tabular}
\end{table*}
%
\begin{table*}
\setlength{\tabcolsep}{2pt}
\center
\caption{Distribution of the residuals. $N$ denotes the number of stars.}
\begin{tabular}{lcc}
\hline
   $\Delta M_g$-interval & $<\Delta M_g>$  & N \\
    \hline
   $\lbrack-0.15,-0.10\rbrack$ & -0.140 &  4 \\
   $\lbrack-0.10,-0.05\rbrack$ & -0.078 &  5 \\
   $\lbrack-0.05, 0.00\rbrack$ & -0.020 & 20 \\
   $\lbrack0.00, 0.05\rbrack$  &  0.024 &  9 \\
   $\lbrack0.05, 0.10\rbrack$  &  0.072 & 11 \\
   $\lbrack0.10, 0.15\rbrack$  &  0.115 &  2 \\
\hline
\end{tabular}
\end{table*}

\begin{table*}
\setlength{\tabcolsep}{5pt}
\center
{\tiny
\caption{Data for the cluster M5 and the field stars used for the application of the procedure with 2MASS.} 
\begin{tabular}{ccccccccc}
\hline
\multicolumn{5}{c}{M5} & & & &  \\
\hline
  $V$ &  $V-J$     &        $J$ &  $(V-J)_0$ &     $J_0$ & & & &  \\
(mag) &  (mag)     &      (mag) &      (mag) &     (mag) & & & &  \\
\hline
21.95 &       2.01 &      19.94 &       1.940 &      19.914 & & & &\\
21.84 &       1.96 &      19.88 &       1.890 &      19.854 & & & &\\
21.60 &       1.84 &      19.76 &       1.770 &      19.734 & & & &\\
21.37 &       1.74 &      19.63 &       1.670 &      19.604 & & & &\\
21.16 &       1.65 &      19.51 &       1.580 &      19.484 & & & &\\
\hline
\end{tabular}
}  
\end{table*}

\begin{table*}
\setlength{\tabcolsep}{5pt}
\center
{\tiny
\caption{Absolute magnitudes $(M_J)_{ev}$ and residuals $\Delta M$ estimated by the procedure explained in our work. 
$(M_J)_{cl}$ and $(M_J)_{\pi}$ denote the absolute magnitudes evaluated by means of the colour-magnitude diagram of 
the M5 cluster and the parallax of the field star, respectively.} 
\begin{tabular}{cccccccccc}
\hline
\multicolumn{10}{c}{Table 18a}\\
\hline
      (1) & (2) &  (3) &   (4) & (1) & (2) &  (3) &   (4) &  & \\
      $(V-J)_0$ & $(M_J)_{cl}$ &  $(M_J)_{ev}$ &   $\Delta M$ &$(V-J)_0$ & $(M_J)_{cl}$ &  $(M_J)_{ev}$ &   $\Delta M$ &  & \\
      (mag) & (mag) &  (mag) &   (mag) & (mag) & (mag) &  (mag) &   (mag) &  & \\
\hline
      0.95 &       3.11 &       3.20 &      -0.09 &       1.37 &       4.74 &       4.68 &       0.06 &  $-$ &$-$\\
      0.97 &       3.27 &       3.34 &      -0.08 &       1.40 &       4.80 &       4.73 &       0.07 &  $-$ &$-$\\
      1.00 &       3.48 &       3.52 &      -0.04 &       1.43 &       4.86 &       4.79 &       0.06 &  $-$ &$-$\\
      1.03 &       3.66 &       3.67 &      -0.01 &       1.45 &       4.89 &       4.83 &       0.06 &  $-$ &$-$\\
\hline
\end{tabular}
}  
\end{table*}
\clearpage

\begin{table*}
\setlength{\tabcolsep}{3pt}
\center
\caption{Distribution of the residuals. $N$ denotes the number of stars.} 
\begin{tabular}{ccc}
\hline
$\Delta M_J$-interval &    $<\Delta M_J>$ &       $N$ \\
     (mag) &           (mag) &        \\
\hline
$(-0.3, -0.2]$&      -0.29 &          1 \\
$(-0.2, -0.1]$&      -0.15 &          8 \\
$(-0.1, 0.0]$ &      -0.03 &         16 \\
$(0.0,  0.1]$ &       0.06 &         36 \\
$(0.1,  0.2]$ &       0.16 &          9 \\
$(0.2,  0.3]$ &       0.25 &          5 \\
$(0.3,  0.4]$ &       0.35 &          4 \\ 
\hline
\end{tabular}
\end{table*}  

\begin{table*}
\setlength{\tabcolsep}{2pt}
\center
\caption{Data for two young clusters.}
\begin{tabular}{cccccc}
\hline
    Cluster & $E(B-V)$ & $\mu_0$ & $[Fe/H]$ & $t$ & Ref \\
              & (mag)  & (mag)   & (dex)    &(Gyr)&     \\
\hline
    NGC 3680 & 0.058 & 9.89  & -0.08 & 1.75  & 1, 2 \\
    NGC 2158 & 0.440 &12.80  & -0.25 & 2.00  & 3, 4 \\
\hline
    \end{tabular}\\
(1) \cite{Cummings12}, (2) \cite{Eggen69}, (3) \cite{Smolinski11}, (4) \cite{Saad01}.\\ 
   \end{table*}
 
\begin{figure*}
\begin{center}
\includegraphics[scale=0.60, angle=0]{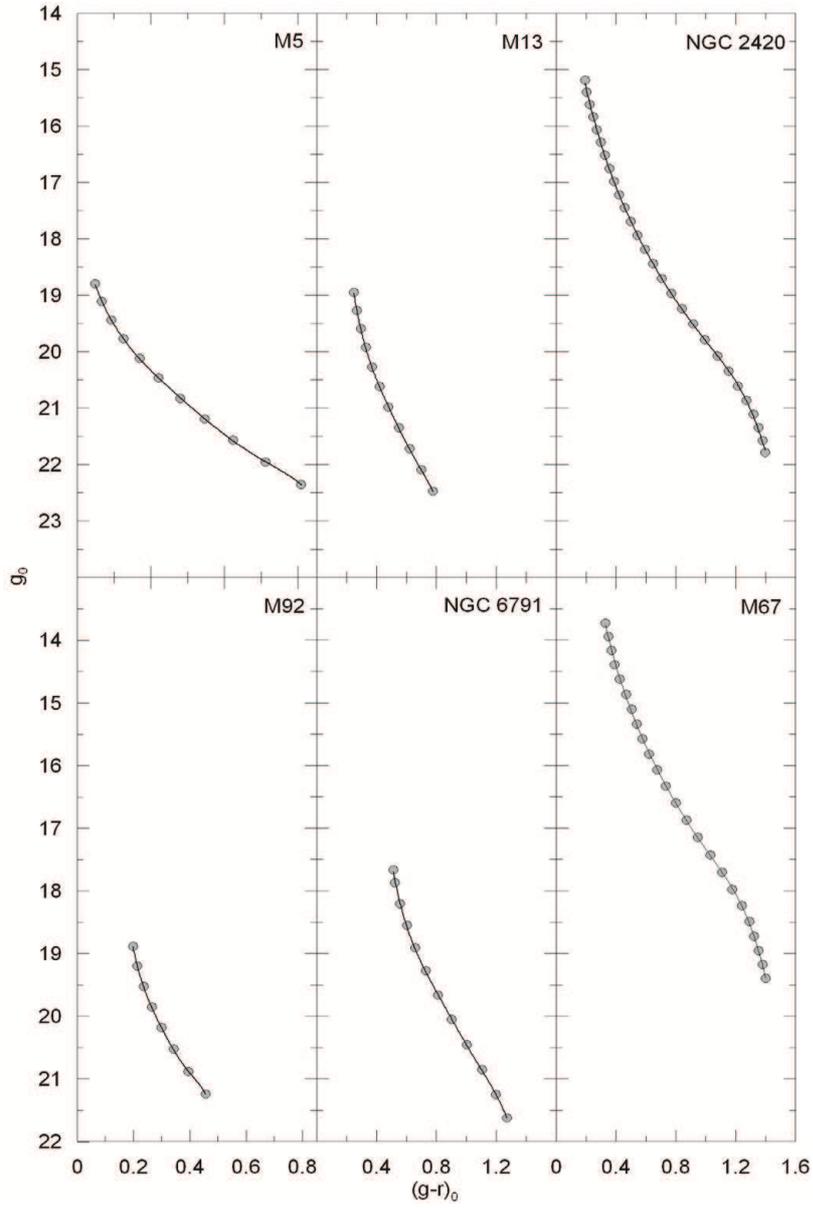}
\caption[] {$g_0\times(g-r)_0$ colour-apparent magnitude diagrams of six stellar clusters 
used for the absolute magnitude calibration with SDSS.} 
\end{center}
\end {figure*}

\begin{figure*}
\begin{center}
\includegraphics[scale=0.60, angle=0]{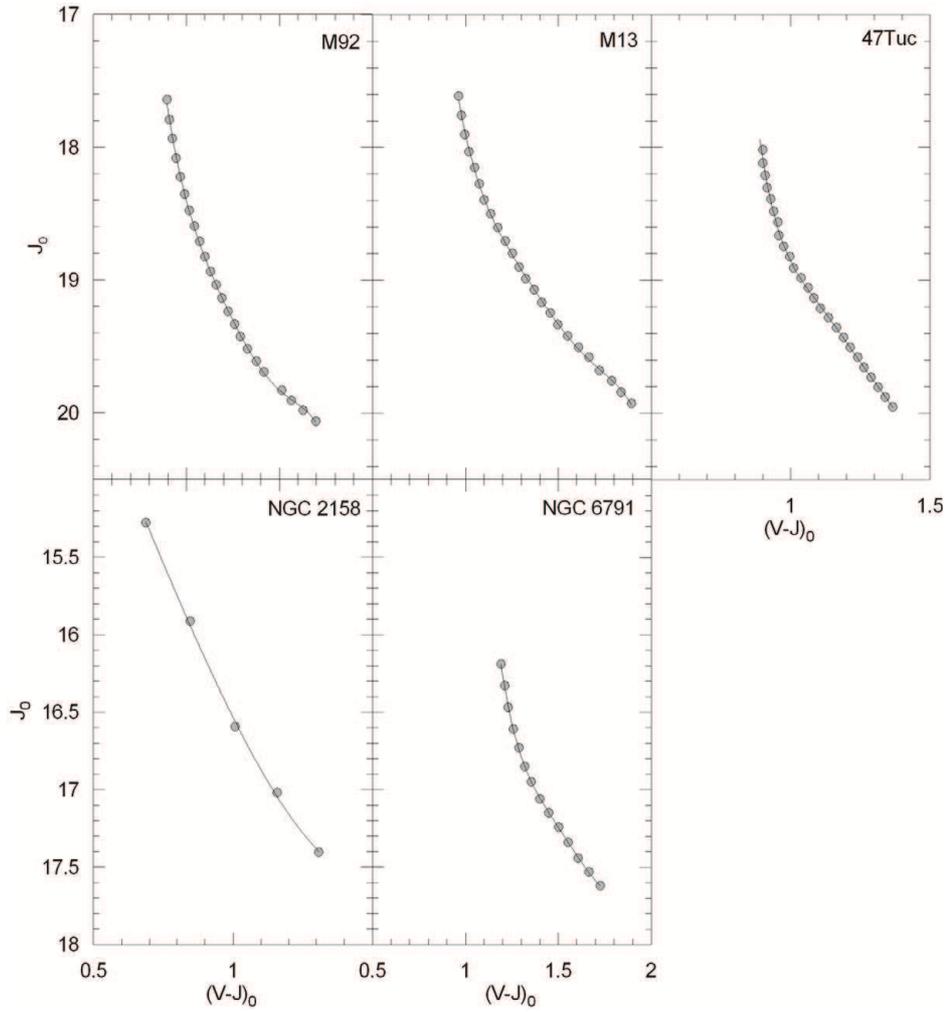}
\caption[] {$J_0\times (V-J)_0$ colour-apparent magnitude diagrams for five clusters used for the $M_J$ absolute magnitude calibration.}
\end{center}
\end{figure*}
 
\begin{figure*}
\begin{center}
\includegraphics[scale=0.60, angle=0]{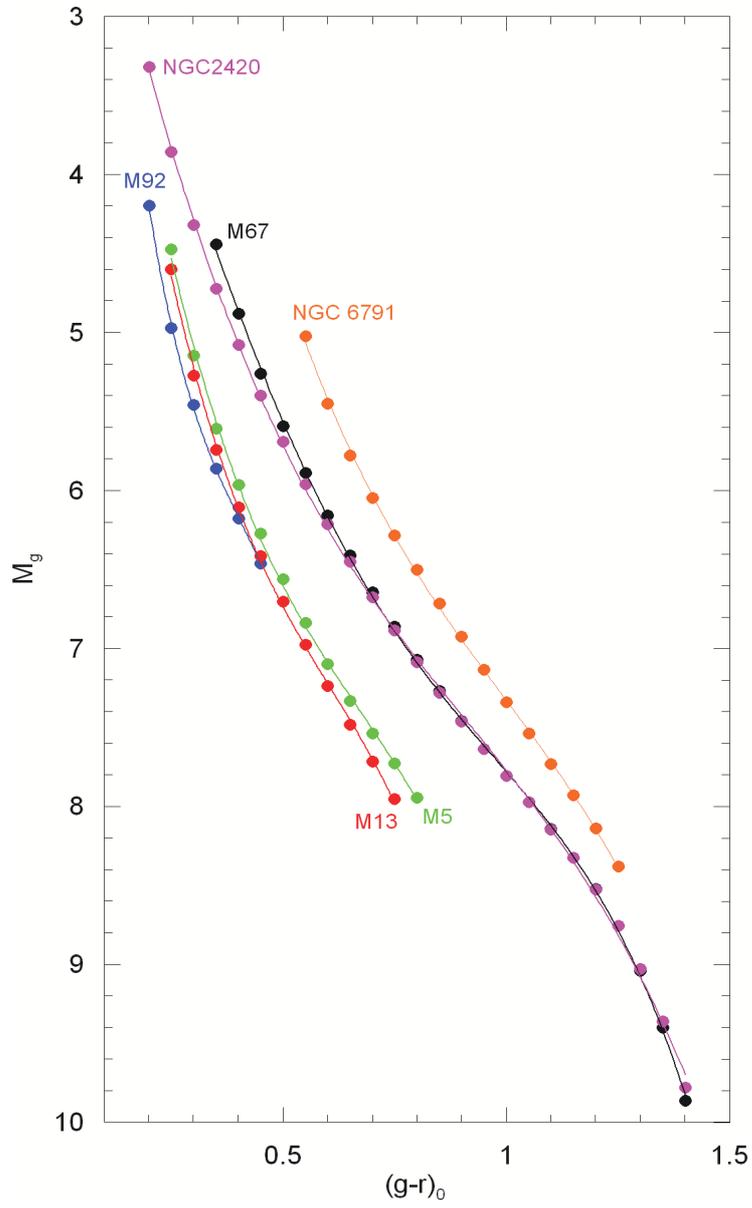}
\caption[] {$M_g\times(g-r)_0$ colour-magnitude diagram for six clusters used for the absolute magnitude calibration with SDSS.} 
\end{center}
\end {figure*}

\begin{figure*}
\begin{center}
\includegraphics[scale=0.50, angle=0]{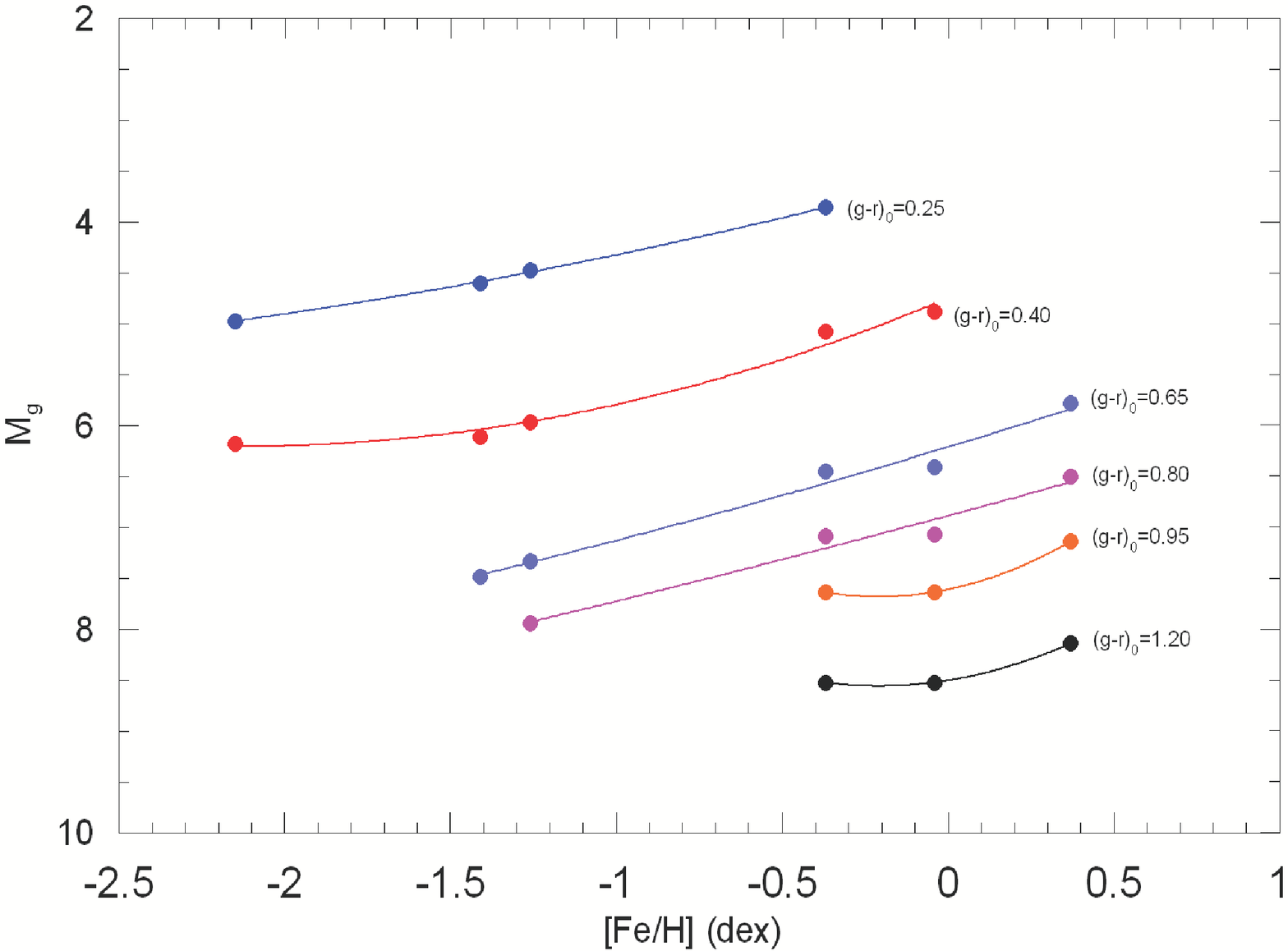}
\caption[] {Calibration of the absolute magnitude $M_g$ as a function of metallicity $[Fe/H]$ for six colour indices.} 
\end{center}
\end {figure*}

\begin{figure*}
\begin{center}
\includegraphics[scale=0.35, angle=0]{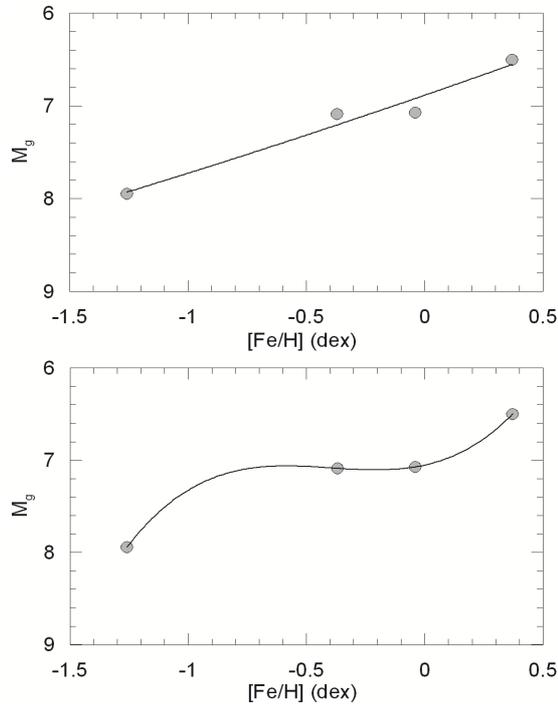}
\caption[]{Comparison the trends of the absolute magnitude calibrations with polynomial degrees $n=2$ (upper panel) and $n=3$ (lower panel) for the colour index $(g-r)_0=0.80$ mag.}
\end{center}
\end {figure*}
 
\begin{figure*}
\begin{center}
\includegraphics[scale=0.50, angle=0]{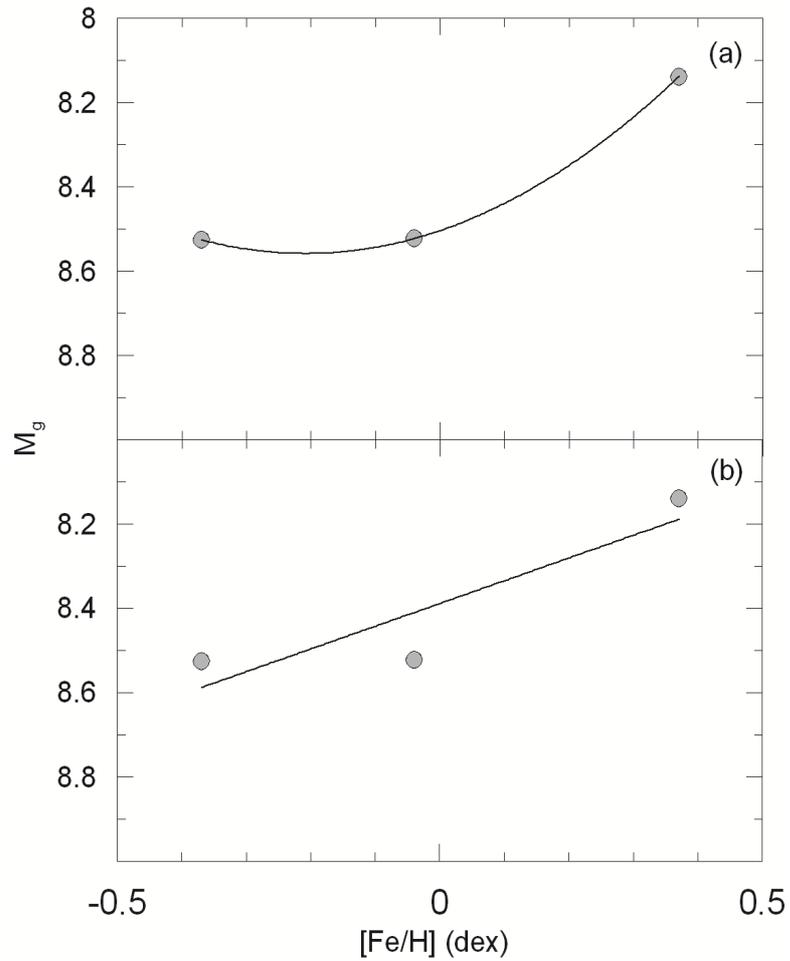}
\caption[] {Two fittings to the metallicities and absolute magnitudes for the colour index $(g-r)_0=1.20$ dex. (a) a quadratic polynomial to three couples and (b) a linear polynomial to three couples.}
\end{center}
\end {figure*}

\begin{figure*}
\begin{center}
\includegraphics[scale=0.60, angle=0]{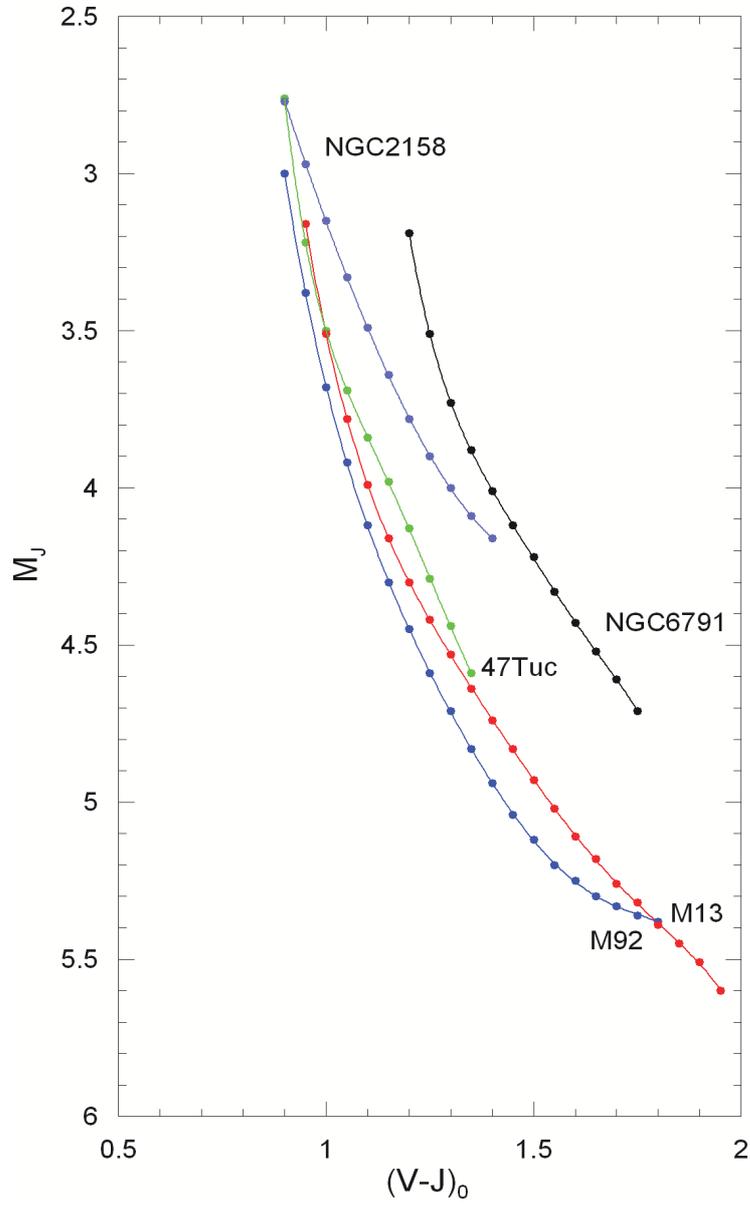}
\caption[]{$M_J\times (V-J)_0$ colour-absolute magnitude diagrams for five clusters used for the absolute magnitude calibration with 2MASS.}
\end{center}
\end{figure*}

\begin{figure*}
\begin{center}
\includegraphics[scale=0.50, angle=0]{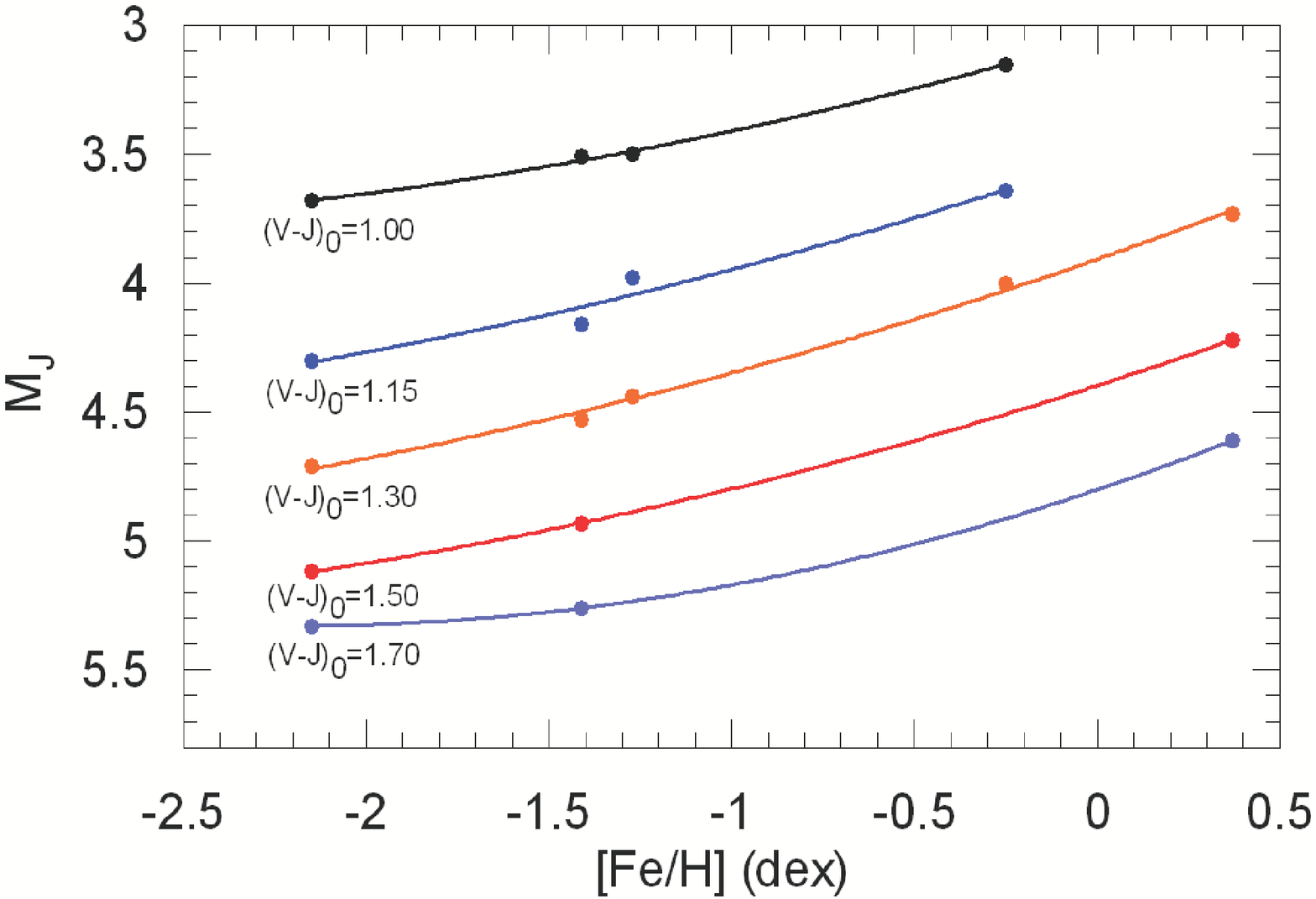}
\caption[] {Calibration of the absolute magnitude $M_J$ as a function of metallicity $[Fe/H]$ for five colour indices.}
\end{center}
\end{figure*}

\begin{figure*}
\begin{center}
\includegraphics[scale=0.40, angle=0]{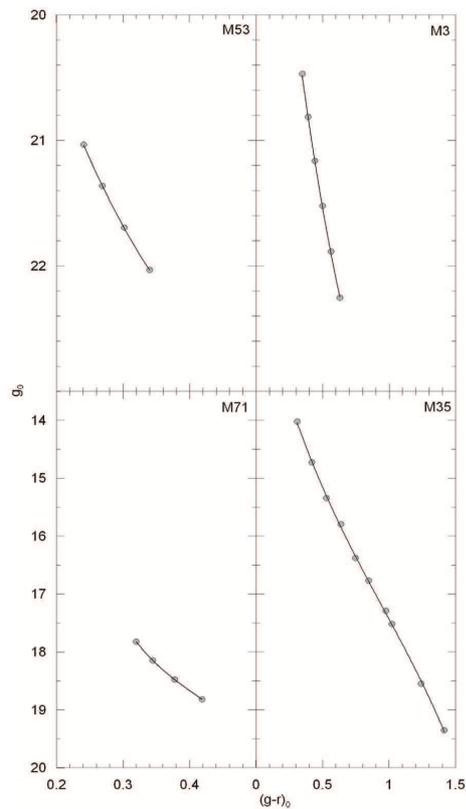}
\caption[] {$g_0\times(g-r)_0$ colour-apparent magnitude diagrams for the Galactic clusters used for the application of the procedure.}
\end{center}
\end {figure*}

\begin{figure*}
\begin{center}
\includegraphics[scale=0.40, angle=0]{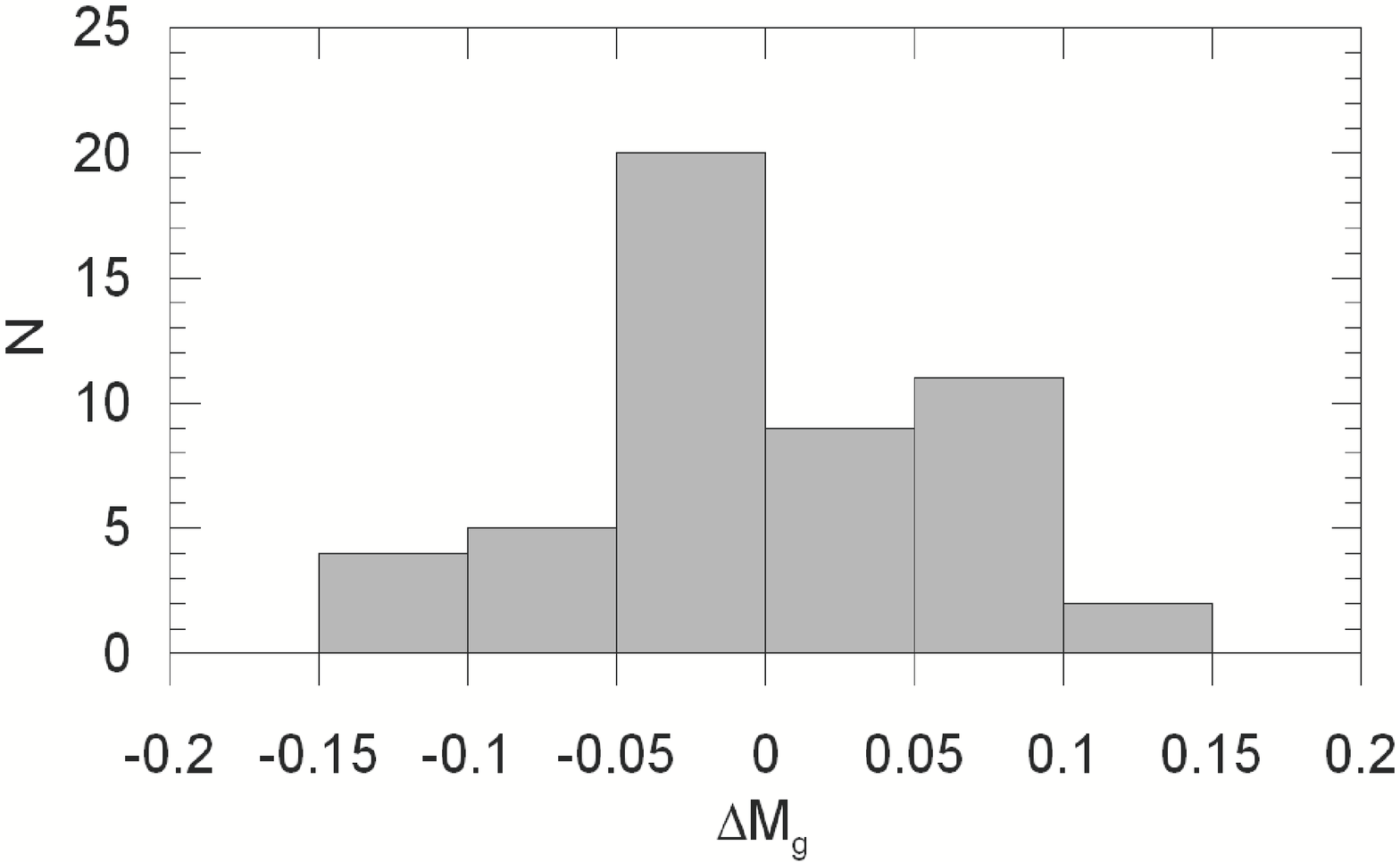}
\caption[] {Histogram of the residuals.} 
\end{center}
\end {figure*}

\begin{figure*}
\begin{center}
\includegraphics[scale=0.40, angle=0]{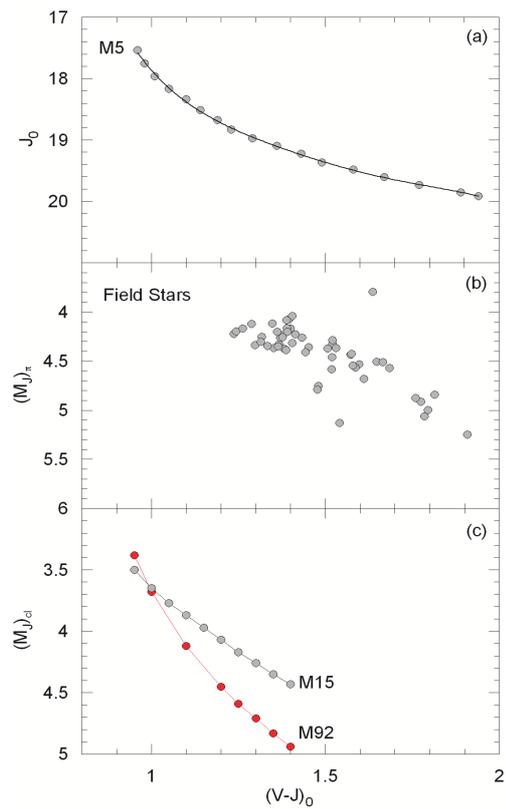} 
\caption[] {$J_0\times (V-J)_0$ colour-apparent magnitude diagram of the cluster M5 (panel a), distribution of the field stars with solar metallicity in the $(V-J)_0$ colour versus ($M_J)_\pi$ absolute magnitude diagram (panel b), and deviation of the $(V-J)_0$ colour $(M_J)_{cl}$ absolute magnitude diagram of the cluster M15 (brighter magnitudes) from the one of M92 (fainter magnitudes).}
\end{center}
\end{figure*}

\begin{figure*}
\begin{center}
\includegraphics[scale=0.45, angle=0]{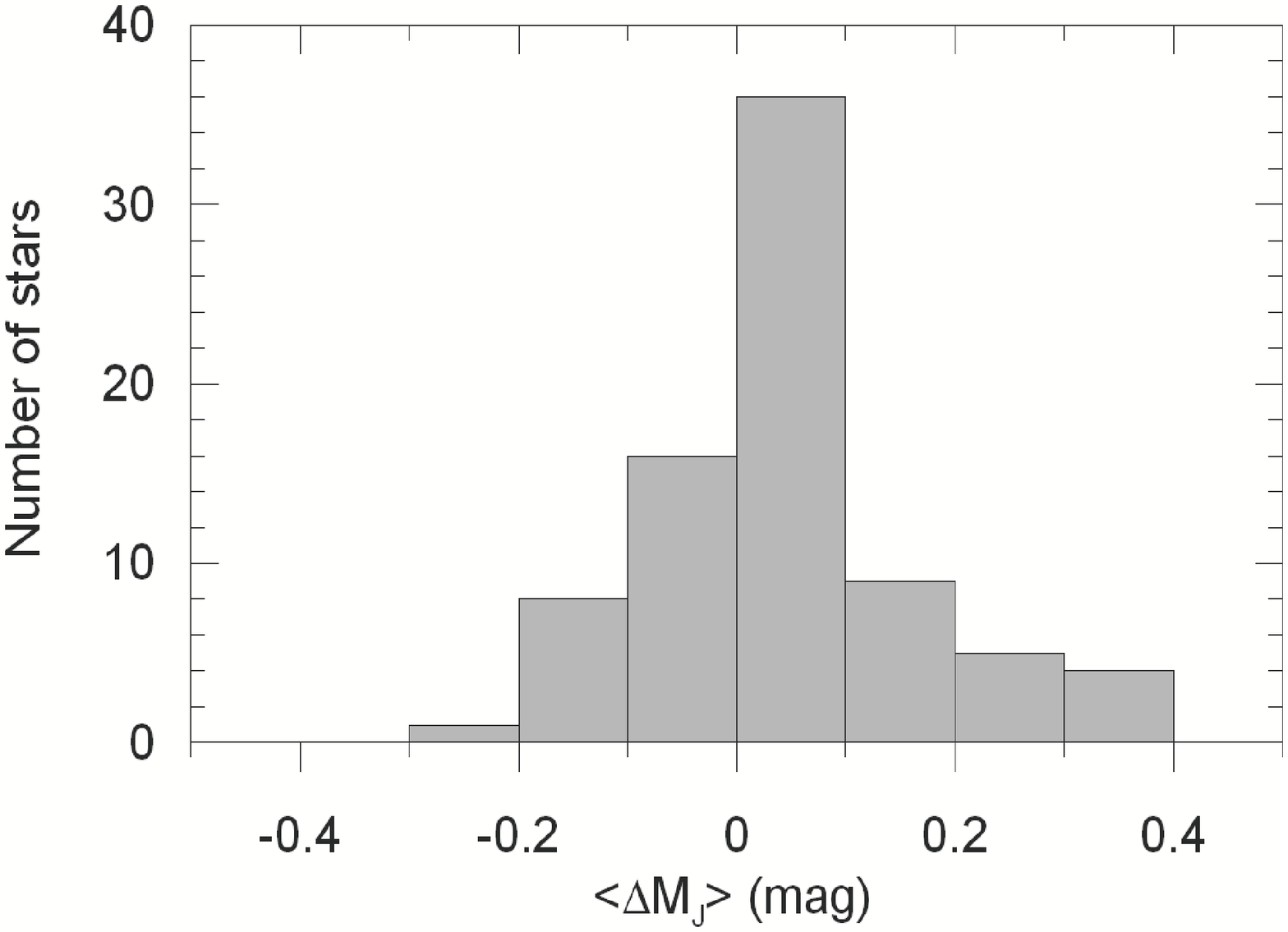}
\caption[] {Histogram of the residuals.}
\end{center}
\end{figure*}

\begin{figure*}
\begin{center}
\includegraphics[scale=0.45, angle=0]{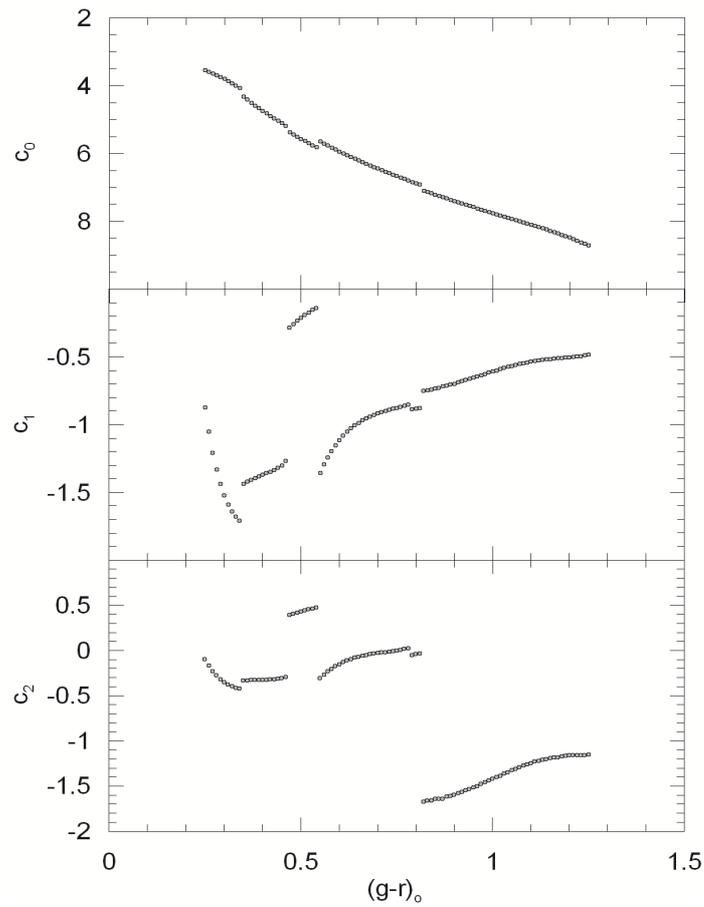}
\caption[] {Variation of $c_0$, $c_1$ and $c_2$ coefficients with colour index $(g-r)_0$ in three panels.} 
\end{center}
\end {figure*}

\end{document}